\documentclass{article}
\usepackage[utf8]{inputenc}
\usepackage[left=1cm, right=1cm, top=2cm, bottom=2cm]{geometry}
\usepackage{changepage}
\usepackage[superscript,biblabel]{cite}
\usepackage{hyperref}
\usepackage{booktabs}
\usepackage{caption}
 \usepackage[super,comma]{natbib}
\usepackage{epsfig,amsfonts,amsmath,graphicx,bm,url}
\usepackage{mathtools,eucal}
\usepackage{xr}
\usepackage{color}
\usepackage{subfig}
\usepackage{multirow}
\usepackage{tikz}               
\usetikzlibrary{shapes.geometric, arrows, shadows, fadings}
\tikzstyle{startstop} = [rectangle, rounded corners, minimum width=3.5cm, minimum height=1.2cm, text centered, draw=black, fill=gray!20]
\tikzstyle{arrow} = [thick,->,>=stealth]

\title{\textbf{Bayesian survival analysis with INLA}}

\author{\textbf{Danilo Alvares$^{*1}$, Janet van Niekerk$^2$, Elias Teixeira Krainski$^2$,}\\ \textbf{H{\aa}vard Rue$^2$ and Denis Rustand$^2$}\\ \ \\
\small $^1$MRC Biostatistics Unit, University of Cambridge, UK \hfill\\
\small $^2$Statistics Program, Computer, Electrical and Mathematical Sciences and Engineering Division, \hfill\\
\small King Abdullah University of Science and Technology (KAUST),\\
\small Thuwal 23955-6900, Kingdom of Saudi Arabia\\}

\date{}

\begin{document}

\maketitle

\abstract{\begin{adjustwidth}{40pt}{40pt}
\normalsize This tutorial shows how various Bayesian survival models can be fitted using the integrated nested Laplace approximation in a clear, legible, and comprehensible manner using the \texttt{INLA} and \texttt{INLAjoint} R-packages. Such models include accelerated failure time, proportional hazards, mixture cure, competing risks, multi-state, frailty, and joint models of longitudinal and survival data, originally presented in the article ``{\it Bayesian survival analysis with BUGS}''.\cite{alvares2021} In addition, we illustrate the implementation of a new joint model for a longitudinal semicontinuous marker, recurrent events, and a terminal event. Our proposal aims to provide the reader with syntax examples for implementing survival models using a fast and accurate approximate Bayesian inferential approach.\ \\\end{adjustwidth} \begin{adjustwidth}{20pt}{120pt}
$^*$\textbf{Correspondence:} Danilo Alvares, MRC Biostatistics Unit,\\ University of Cambridge, CB20SR, UK. Email: danilo.alvares@mrc-bsu.cam.ac.uk\end{adjustwidth}}

\normalsize 
\section{Introduction} \label{sec:intro}

Time until an event of interest is a very common data type in many fields such as medicine (e.g., time to death), biology (e.g., resistance time of a bacterium), political science (time until a protest occurs), engineering (e.g., failure time of a compress), and so on. To model this type of problem, some features must be considered: (i) non-negative variable of interest (and potentially positive skewness), (ii) censoring, and (iii) truncation.\cite{klein2003, collet2015}

Censoring is a type of incomplete/partial information as the exact time of the event of interest is not recorded during the study time window. For instance, an individual has already experienced the event of interest before starting the study (left-censoring), or the study ends without the subject having experienced such an event (right-censoring), or the event occurred within an interval instead of being observed exactly (interval-censoring). Truncation occurs when only those individuals whose event time lies within a certain interval are observed. Specifically, truncation is due to sampling bias in the study design and can be classified as left, right, or interval truncation. For instance, there are studies that define a minimum age (left-truncation) to consider individuals at risk (e.g., 65 years of age for Alzheimer's patients), so any record of individuals younger than the minimum age is not observed/considered and all survival/risk analyses are conditioned on the minimum age.  See Kleinbaum and Klein\cite{kleinbaum2012} for more examples of censoring and truncation.

The distinct features of time-to-event data are what make the use of standard statistical techniques unfeasible. However, this problem is circumvented with survival (or reliability) theory, which proposes mechanisms that incorporate such features into the inferential process.\cite{aalen2008} To introduce these mechanisms, let $T$ be a non-negative random variable representing the time-to-event with hazard and survival functions expressed as $h(t \mid {\bm \theta})$ and $S(t \mid {\bm \theta})$, respectively, where ${\bm \theta}$ is a parameter vector. The contribution of a certain individual to the likelihood function depends on their type of censoring and whether the study presents truncation. In order to explore all scenarios, let's define some elements: $t$ is the exact/observed survival time; $C_{l}$ is the left-censoring time, indicating that the event occurred before this time (e.g., before the data is collected or study has started); $C_{r}$ is the last time of registration of an individual (loss-to-follow-up censoring) or end of the study period (administrative censoring) and the event of interest did not occur (right-censoring); $L$ is the lower limit of time for individuals to be at risk (left-truncation); and $R$ is the upper limit of time for individuals to be at risk (right-truncation). Table \ref{table:censtrunc} summarises how the peculiarities of survival data influence an individual's contribution to the likelihood function.

\begin{center}
\begin{table}[htb] \centering
\caption{Types of censoring and truncation and their influences on the likelihood function. \label{table:censtrunc}}
\begin{tabular*}{500pt}{@{\extracolsep\fill}ccc@{\extracolsep\fill}}
\cmidrule[\heavyrulewidth]{2-3}
 & \multicolumn{2}{c}{Likelihood = numerator/denominator}  \\
\cline{2-3}
 &    Censoring         &   Truncation  \\
 &   (numerator)        &   (denominator) \\
\midrule
No        &    $h(t \mid {\bm \theta})S(t \mid {\bm \theta})$      &   1  \\
Left      &    $1-S(C_l \mid \bm \theta)$                          &   $S(L \mid \bm \theta)$  \\
Right     &    $S(C_r \mid \bm \theta)$                            &   $1-S(R \mid \bm \theta)$  \\
Interval  &    $S(C_l \mid \bm \theta) - S(C_r \mid \bm \theta)$   &   $S(L \mid \bm \theta) - S(R \mid \bm \theta)$  \\
\bottomrule
\end{tabular*}
\end{table}
\end{center}

In the last decades, many books, articles, and packages have proposed methodologies for the survival framework. In particular, the Bayesian approach has been widely used to measure uncertainties, incorporate prior knowledge, and make dynamic predictions.\cite{sinha1997, ibrahim2001, martino2011, rizopoulos2014, bartos2022} However, Markov chain Monte Carlo (MCMC) methods can be quite time-consuming for complex survival models and/or massive data. Integrated nested Laplace approximation (INLA) is a convenient alternative to MCMC for Bayesian inference due to its execution speed and accuracy.\cite{rue2009} It can be used to fit most survival models\cite{martino2011} but there is limited literature highlighting it, mostly because INLA was further considered for survival analysis recently\cite{niekerk2021,niekerk2021cr} and its use through the \texttt{INLA} R-package becomes tricky when fitting survival models more complex than a simple parametric proportional hazards model.\cite{gomezrubio2020}

Inspired by ``{\it Bayesian survival analysis with BUGS}'',\cite{alvares2021} this tutorial aims to implement the same survival models presented in the original work, and more complex models, using INLA.

The rest of the work is organised as follows. Section \ref{sec:inla} introduces the INLA methodology and its R-packages that will support the implementation of our Bayesian survival models. Sections \ref{sec:AFT} to \ref{sec:JM1} describe accelerated failure time, proportional hazards, mixture cure, competing risks, multi-state, frailty, and joint models of longitudinal and survival data, respectively, as in Alvares et al\cite{alvares2021}. Section \ref{sec:JM2} exemplifies the implementation of joint models of longitudinal semicontinuous, recurrent events and terminal event data using INLA. Section \ref{sec:PS} gives some insights on model check and illustrates how to perform prior sensitivity analysis. Section \ref{sec:CMP} compares computational time and posterior distributions between INLA and MCMC for the models from Sections \ref{sec:AFT} to \ref{sec:JM1}. Finally, the article is concluded by a brief discussion in Section \ref{sec:conclusions}. R codes are available at \url{https://github.com/DenisRustand/Bayes-surv-INLA}.

\section{The integrated nested Laplace approximation (INLA)} \label{sec:inla}

The integrated nested Laplace approximation (INLA) is a deterministic alternative to sampling-based methods.\cite{rue2009} It performs fast and accurate Bayesian inference for latent Gaussian models (LGM) using Laplace approximations and efficient numerical integrations. Recent developments within the INLA methodology, using variational Bayes with a Laplace approximation,\cite{niekerk2022} further improve accuracy and efficiency, making INLA an attractive approximate Bayesian inference method for fast and accurate results while maintaining the flexibility of the models being considered.

Consider the data $\bm y$, latent field $\bm x$ and a set of hyperparameters $\bm\theta$. A statistical model can be formulated as an LGM if it is a Bayesian hierarchical model where each data point is connected to one linear predictor, such that the data is conditionally independent given the latent field and the set of hyperparameters (i.e., $\left(y_i \mid \bm x, \bm\theta\right) \perp \left(y_j \mid \bm x, \bm\theta\right)$\.). An LGM can thus be defined as follows:  
\begin{eqnarray*}
&&y_i \sim f(\eta_i, \bm\theta),  \quad \quad \bm\eta = \bm A\bm x \quad \text{with likelihood function }\quad  \prod_{i=1}^nf(y_i \mid \bm x,\bm \theta)\\
&&\bm x \mid \bm\theta \sim \text{Normal}(\bm 0, \, \bm Q^{-1}_{\text{prior}})\\
&&\bm\theta \sim g(\cdot)
\end{eqnarray*}

Hence, we have three levels in the model, the likelihood from the data (this can be constructed by many different models like Binomial, Poisson, Gamma, Beta, Student's t, etc), the latent field (containing all the fixed effects and random effects) is assumed to be a Gaussian Markov random field (GMRF), $p(\bm x \mid \bm\theta)$, and the set of hyperparameters (containing hyperparameters from both the likelihood model and the GMRF) can assume any prior, $g(\bm\theta)$. The Gaussian assumption is thus only for the latent field prior.

The INLA methodology can be summarised in mainly two stages, in the first stage the posterior density of the hyperparameters, $\pi(\bm \theta \mid \bm y)$, is approximated while the posterior density of the latent field, $\pi(\bm x \mid \bm y)$,  is approximated in the second stage. Various numerical and computational techniques are used in both stages, including the use of a smart gradient\cite{fattah2022} in the optimisation step of stage 1, an optimisation-based low-rank implicit variational Bayes correction\cite{niekerk2021bis} in stage 2, and others - more details are available in works dedicated to the INLA methodology.\cite{rue2009, rue2017, niekerk2022} The various stages can be summarised as follows:
\begin{itemize}
    \item Stage 1:  
Define 
\begin{equation*}
    \pi(\bm\theta \mid \bm y) = \frac{p(\bm x \mid \bm\theta)g(\bm\theta)\prod_{i=1}^nf(y_i \mid \bm x, \bm\theta)}{\pi(\bm x \mid \bm y, \bm\theta)}.
\end{equation*}
Now we approximate $\pi(\bm\theta \mid \bm y)$ with $\tilde{\pi}(\bm\theta \mid \bm y)$ using a Gaussian approximation of $\pi(\bm x \mid \bm y, \bm\theta)$ around its mode $\bm\mu$ such that
\begin{equation*}
    \tilde{\pi}(\bm\theta \mid \bm y) = \left.\frac{p(\bm x \mid \bm\theta)g(\bm\theta)\prod_{i=1}^nf(y_i \mid \bm x, \bm\theta)}{\pi_G(\bm x \mid \bm y, \bm\theta)}\right|_{\bm x = \bm\mu},
\end{equation*}
with posterior marginals 
\begin{equation*}
    \tilde{\pi}(\theta_j \mid \bm y) = \int \tilde{\pi}(\bm\theta \mid \bm y)d\bm\theta_{-j}.
\end{equation*}

    \item Stage 2: Using variational Bayes, correct the mean of the Gaussian approximation $\pi(\bm x \mid \bm y, \bm\theta)$ to obtain a new improved joint mean $\bm\mu^*$,\cite{niekerk2021bis} from which the conditional posteriors $\pi(x_j \mid \bm y, \bm\theta)$ can be extracted. Then the posterior marginals of the latent field can be calculated as
    \begin{equation*}
    \tilde{\pi}(x_j \mid \bm y) = \int \pi(x_j \mid \bm y,\bm\theta)\tilde{\pi}(\bm\theta \mid \bm y)d\bm\theta.
\end{equation*}
    Another strategy (originally proposed in Rue et al\cite{rue2009}) uses subsequent Gaussian approximations (equal to the size of the latent field, which can be large) to find the element-wise conditional posteriors, and is named the Laplace strategy. The use of the low-rank Variational Bayes correction achieves the same accuracy with much less computational cost as the Laplace strategy. 
\end{itemize}

The integration over the hyperparameter space is done using a central composite design (CCD) approach where integration points and corresponding weights are used for the numerical integration. If the dimension of the hyperparameter space is large (roughly more than $20$), this integration can be slow, and either an empirical Bayes approach can be used in INLA, or parallelisation.\cite{gaedke2022} The empirical Bayes approach uses only the mode of the hyperparameter posterior and circumvents the integration in totality. This method presents excellent frequentist properties for multivariate joint longitudinal-survival models although it does not fully account for uncertainty.\cite{rustand2022}

\subsection{Survival analysis with INLA} \label{subsec:survinla}

At first glance, it might seem that survival models are not LGMs, but previous works show how various survival models can be formulated as LGMs and thus INLA can be used for approximate Bayesian inference of these models.\cite{martino2011, niekerk2019, niekerk2021cr}  The most common method for analysing survival data assumes proportional hazards, initially proposed by Sir David Cox \citep{Cox72}. It has been shown that the Cox model can be approximated using a Poisson regression framework. \citep{Holford76, Whitehead80, Johansen83} The approximation involves transforming the continuous-time hazard function into a discrete-time event count intensity function. By considering sufficiently small time intervals, the hazard can be approximated as relatively constant within each interval. Consequently, the number of events in each interval follows a Poisson distribution, with the mean proportional to the hazard in that specific interval. This approximation becomes more accurate as the time intervals become narrower.

The hazard function of the Cox model is given by $h(t\mid\bm\theta)$ and the Poisson approximation in discrete time is based on the assumption that the number of events $N_i$ in a small time interval of length $\Delta t_i$ follows a Poisson distribution:
$$N_i \sim \text{Poisson}(\lambda_i),$$
where the mean $\lambda_i$ is proportional to the hazard within the interval such that:
$$\lambda_i = h\left(t_i\mid\bm\theta\right) \Delta t_i.$$

The Poisson regression model aligns with the LGM framework which implies that INLA can be used as a unified framework for approximate Bayesian inference of various complex survival models, some of which are considered here.

Since this approach involves a decomposition of the follow-up into many small intervals, we can also take advantage of this data augmentation to account for time-dependent components in the hazard function, thus allowing to fit joint models for longitudinal and survival outcomes with time-dependent association structures. Parametric proportional hazards models and accelerated failure time models can also be integrated into this framework by specifying a specific distribution for the baseline hazard approximation (e.g., exponential and Weibull). Various censoring schemes, such as right, left, or interval censoring, are also accommodated. 

For a semi-parametric approach, the default option for baseline hazards in \texttt{INLAjoint} is a random walk model. In most cases, one does not know much about the shape of the baseline hazard and thus could let the data determine it. However, it is important to avoid overfitting, this is why we use Bayesian smooth splines (i.e., random walk of order 1 or 2) to avoid abrupt changes in the baseline hazard which is usually unrealistic. The random walks are defined by hyperparameters that are part of the model parameters estimated by \texttt{INLAjoint}. The baseline hazard in \texttt{INLAjoint} is specified with the argument ``basRisk'' with the default value being ``rw1'' (i.e., random walk of order 1) and alternative options being ``rw2'' (i.e., random walk of order 2), ``exponentialsurv'' and ``weibullsurv'' (exponential and Weibull parametric distributions, respectively).

\subsection{The INLA and INLAjoint R-packages} \label{subsec:inlajoint}
The INLA approach is available for R users in the \texttt{INLA} R-package, sometimes termed R-INLA, and can be installed from its website (\url{https://www.r-inla.org/download-install}). The package is ever-evolving with new functionality being added continuously. The core method is implemented in C language considering the state of the art of the necessary numerical algorithms. It uses two levels of parallelisation to run simultaneously multiple tasks on shared memory with OPENMP and each one can also run in parallel. This parallel computation approach enable the use of R-INLA for complex models and big data applications.\cite{niekerk2019, gaedke2022}

Because the \texttt{INLA} R-package is able to fit a wide variety of statistical models (e.g., generalised linear and additive mixed effects models, spatial and spatio-temporal models, temporal models, stochastic volatility models, disease mapping models, survival models),\cite{rue2017} it is required to formulate the survival model as an LGM, which becomes cumbersome for complex models. In this context, the \texttt{INLAjoint} R-package is an interface that facilitates the use of \texttt{INLA} for the more specific sub-class of models that involves longitudinal and/or survival outcomes. It makes the use of \texttt{INLA} for those models much simpler while being flexible enough to fit most survival models that \texttt{INLA} can fit. In this article, we will use this package for all the survival models fitted as it keeps the syntax simple, especially when considering the default priors.\cite{rustand24}

The R package \texttt{INLAjoint} is available on CRAN (\url{https://cran.r-project.org/web/packages/INLAjoint/index.html}) and Github (\url{https://github.com/DenisRustand/INLAjoint}). It revolves around the main function \texttt{joint()} used to fit the models:
\begin{verbatim}
joint(
  formLong = NULL, dataLong = NULL, id = NULL, timeVar = NULL,
  family = "gaussian", link = "default", corLong = FALSE,
  corRE = TRUE, formSurv = NULL, dataSurv = NULL, 
  basRisk = "rw1", NbasRisk = 15, cutpoints = NULL, 
  assocSurv = NULL, assoc = NULL, control = list(), ...
)
\end{verbatim}

The arguments \texttt{formLong} and \texttt{formSurv} corresponds to the formula for longitudinal and survival models, respectively. Both can be specified as lists in case of multivariate outcomes and can be used simultaneously for joint models of longitudinal and survival data. The arguments \texttt{dataLong} and \texttt{dataSurv} give the datasets corresponding to each formula, although it is possible to specify only one dataset that contains information for all formulas.

When fitting longitudinal models, the following arguments can be specified: \texttt{id} (name of the grouping variable for repeated measurements), \texttt{timeVar} (name of the time variable), \texttt{family} (family of the likelihood, which is specified as a vector in case of multivariate longitudinal outcomes), \texttt{link} (link between the linear predictor and the outcome for each family), \texttt{corLong} (boolean to indicate whether longitudinal outcomes should have their random effects correlated), \texttt{corRE} (boolean indicating if random effects of a given longitudinal model should be correlated or independent, this should be a vector if there are multiple groups of random effects). On the other hand, for survival submodels, the following arguments can be specified: \texttt{basRisk} gives the baseline risk function, with default random walk but it can also be specified as a parametric distribution as explained in Section \ref{subsec:survinla}, \texttt{NbasRisk} gives the number of cutpoints to decompose the follow-up into small intervals as described in Section \ref{subsec:survinla}, this allows to approximate the Cox model with a Poisson distribution and it is also used for the evolution over time of baseline hazards and time-dependent components in the baseline hazards function. Alternatively, the argument \texttt{cutpoints} allows to manually specify the cutpoints instead of the default equidistant cutpoints between baseline and the maximum follow-up time, which should be provided as a vector.

Finally, parameters related to the association between submodels in the context of joint modelling are: \texttt{assocSurv} which allows to share a frailty term (i.e., random effect in a survival submodel) into another survival submodel, as illustrated in our example described in Section \ref{sec:JM2} and \texttt{assoc} which allows to specify the association between longitudinal submodels and survival submodels, with the possibility to share the entire linear predictor (``CV'' for current value association), the individual deviation from the population mean (``SRE'' for shared random effects), the rate of change of the linear predictor (``CS'' for current slope) and the individual deviation defined by each random effect shared and scaled separately (``SRE\_ind'' for shared random effects independently). 

The last argument is \texttt{control}, which is a list of options that can be specified. Details about these available control options are listed in the help page of the joint function that can be accessed with the \texttt{?joint} call in R after loading \texttt{INLAjoint}.

\subsection{Priors in the INLAjoint R-package} \label{subsec:inlajointpriors}

The default prior distributions in \texttt{INLAjoint} are specified in a weakly informative framework. All regression coefficients and the Weibull log-scale parameter follow a Normal($\mu=0, \, \sigma^2=100$) prior. For hyperparameters, some default priors are penalising complexity priors as defined by Simpson et al\cite{simpson2017penalising}. PC priors are derived based on an exponential prior on the distance between a model and its simpler counterpart, obtained by a specific value or limit of the hyperparameter. The prior contracts toward the simpler model at a fixed exponential rate. These priors are computationally simple and allow for adjustment of the prior informativeness with a single user-defined contraction parameter.
Particularly, the Weibull shape parameter follows a penalised complexity prior, PC($5$) (see \texttt{inla.doc("pcalphaw")} for more details). This type of prior is an informative shrinking alternative to existing prior choices such as Gamma or Uniform distributions, as described in van Niekerk et al\cite{niekerk2021}.  The precision parameter for the random walks also follows a penalised complexity prior such that the probability of the marginal standard deviation of the constrained random walk specified at the knots being above a given threshold $P(\sigma > 0.5) = 0.01$ is fixed (see \texttt{inla.doc("pc.prec")} for more details). A log-Gamma($a=0.01, \, b=0.01$) prior is specified to the frailty term variance (see \texttt{inla.doc("loggamma")} for more details). For joint model specifications, the prior on the multivariate random effects precision matrix is a Wishart($r, \, {\bm \Omega}^{-1}$) where $r=10$ and $\bm \Omega$ is an identity matrix (see \texttt{inla.doc("iidkd")} for more details). 

Obviously, all default prior distributions can be modified according to user preferences. Indeed, in the call of the \texttt{joint} function, priors can be specified through control options, for example, priors for fixed effects can be specified with the \texttt{priorFixed} control argument, which is a list with entries: \texttt{mean}, \texttt{prec}, \texttt{mean.intercept} and \texttt{prec.intercept} where \texttt{mean} and \texttt{prec} are the mean and precision (i.e., inverse of the variance) of the fixed effects, respectively and \texttt{mean.intercept} and \texttt{prec.intercept} are the corresponding parameters for the intercept. Default values are mean 0 and precision 0.01 (i.e., variance 100). Additionally, the control option \texttt{priorRandom} is a list with prior distribution for the multivariate random effects (inverse-Wishart). Default is \texttt{list(r=10, R=1)}, see \texttt{inla.doc("iidkd")} for more details. Prior distributions for baseline hazards hyperparameters (e.g., parameters related to parametric distributions or random walks) can be specified through the \texttt{baselineHyper} control argument. Details about the default and available prior distributions can be displayed with the \texttt{inla.doc} command, for example \texttt{inla.doc("weibullsurv")} for the Weibull baseline hazards priors. While multiple options for each priors are available, there is no limitation for the specification of alternative priors and more distributions can be added upon request. Finally, the function \texttt{priors.used} applied to an \texttt{INLAjoint} object returns the priors that were used to fit the model.
\section{Accelerated failure time models} \label{sec:AFT}

\subsection{Motivation: larynx dataset (from the \texttt{KMsurv} R-package)} \label{subsec:aftdata}

The {\it larynx} dataset contains observations of 90 male larynx-cancer patients, diagnosed and treated in the period 1970 to 1978.\cite{kardaun1983} Each patient has the following variables:
\begin{itemize}
	\item \texttt{stage}: disease stage (1--4).
	\item \texttt{time}: time (in months) from first treatment until death, or end of study.
	\item \texttt{age}: age (in years) at diagnosis of larynx cancer.
	\item \texttt{diagyr}: year of diagnosis of larynx cancer.
	\item \texttt{delta}: death indicator (1: if patient died; 0: otherwise).
\end{itemize}

The distribution of disease stages was 33 for stage 1, 17 for stage 2, 27 for stage 3, and 13 for stage 4; the observed survival times had 0.1 years as minimum, 10.7 as maximum, and 4 as median, with 44.4\% right-censored; ages at diagnosis ranged between 41 and 86 years old with mean and median at 64.6 and 65, respectively; and the years of diagnosis fluctuated between 1970 and 1978, where the highest number of cases (19) was observed in 1976.

\subsection{Model specification}

Time to death, $T$, can be modelled by an accelerated failure time (AFT) model as follows:
\begin{equation}
\log(T) = \beta_{1} + \beta_{2}\mbox{I}_{(\texttt{stage}=2)} + \beta_{3}\mbox{I}_{(\texttt{stage}=3)} + \beta_{4}\mbox{I}_{(\texttt{stage}=4)} + \beta_{5}\texttt{age} +\beta_{6}\texttt{diagyr} + \sigma \epsilon, \label{eq:larynx1}
\end{equation}

\noindent where $\beta_{1}$ is an intercept; $\mbox{I}_{(\texttt{stage}=\cdot)}$ is an indicator variable for \texttt{stage}=2, 3, 4 with regression coefficients $\beta_{2}$, $\beta_{3}$, and $\beta_{4}$, respectively (\texttt{stage}=1 is considered as the reference category); and $\beta_{5}$ and $\beta_{6}$ are regression coefficients for \texttt{age} and \texttt{diagyr} covariates, respectively. The errors $\epsilon$'s are independent and identically distributed (i.i.d.) random variables that follow a standard Gumbel distribution and $\sigma$ is a scale parameter. Prior distributions are set according to Section~\ref{subsec:inlajointpriors}.

\subsection{Model implementation} \label{subsec:aftimpl}

We start by loading the data, rescaling the continuous variables \texttt{age} and \texttt{diagyr}, and defining the categorical variable \texttt{stage} as a factor:
\begin{verbatim}
R> library(KMsurv)
R> data(larynx)
R> larynx$age <- as.numeric(scale(larynx$age))
R> larynx$diagyr <- as.numeric(scale(larynx$diagyr))
R> larynx$stage <- as.factor(larynx$stage)
\end{verbatim}

So, we fit a Weibull model with an INLA survival object as the outcome:
\begin{verbatim}
R> m1.weib <- joint(formSurv = inla.surv(time = time, event = delta) ~ stage + age + diagyr,
+                   basRisk = "weibullsurv", dataSurv = larynx, control = list(config=TRUE))
\end{verbatim}
However, specification \eqref{eq:larynx1} is a Gumbel AFT model. The relationship between the regression coefficients of the Weibull and Gumbel AFT models is just a sign inversion, so we can do this transformation with the \texttt{inla.tmarginal} function that allows applying any function to transform the marginals and \texttt{inla.zmarginal} that computes summary statistics of marginal posteriors:
\begin{verbatim}
R> m1.gumb <- sapply(m1.weib$marginals.fixed, function(m) 
+                    inla.zmarginal(inla.tmarginal(function(x) -x, m)))
\end{verbatim}

Rearranging some terms, the posterior distribution summary is given by:
\begin{verbatim}
R> gumb.scale <- inla.tmarginal(function(x) 1/x, m1.weib$marginals.hyperpar[[1]])
R> rbind("Gumbel scale"=inla.zmarginal(gumb.scale), t(m1.gumb))
                mean     sd quant0.025 quant0.5 quant0.975
Gumbel scale  0.9794 0.0266     0.9198   0.9829     1.0210
Intercept_S1  2.5846 0.2718     2.0506   2.5841     3.1180
stage2_S1    -0.1294 0.4594    -1.0323  -0.1302     0.7719
stage3_S1    -0.6580 0.3523    -1.3505  -0.6586     0.0333
stage4_S1    -1.6866 0.4183    -2.5086  -1.6873    -0.8659
age_S1       -0.2125 0.1532    -0.5135  -0.2127     0.0881
diagyr_S1     0.0738 0.1550    -0.2308   0.0736     0.3779
\end{verbatim}

The scale parameter can be obtained by $\sigma=1/\alpha$, where $\alpha$ is the Weibull shape parameter.

The marginal posterior distributions are available in the output of the joint function. It is easy to convert those marginals from the Weibull formulation to the Gumbel AFT model by using the \texttt{inla.tmarginal} function:
\begin{verbatim}
R> m1.marg <- append(list("Gumbel scale" = gumb.scale),
                  lapply(m1.weib$marginals.fixed, function(m) inla.tmarginal(function(x) -x, m)))
R> sapply(1:7, function (x) plot(m1.marg[[x]], type = "l", xlab = "", ylab = "", 
                                 main = paste0("Density of ", names(m1.marg)[x])))
\end{verbatim}

\begin{figure}[ht]
\centering
\includegraphics[scale=0.75]{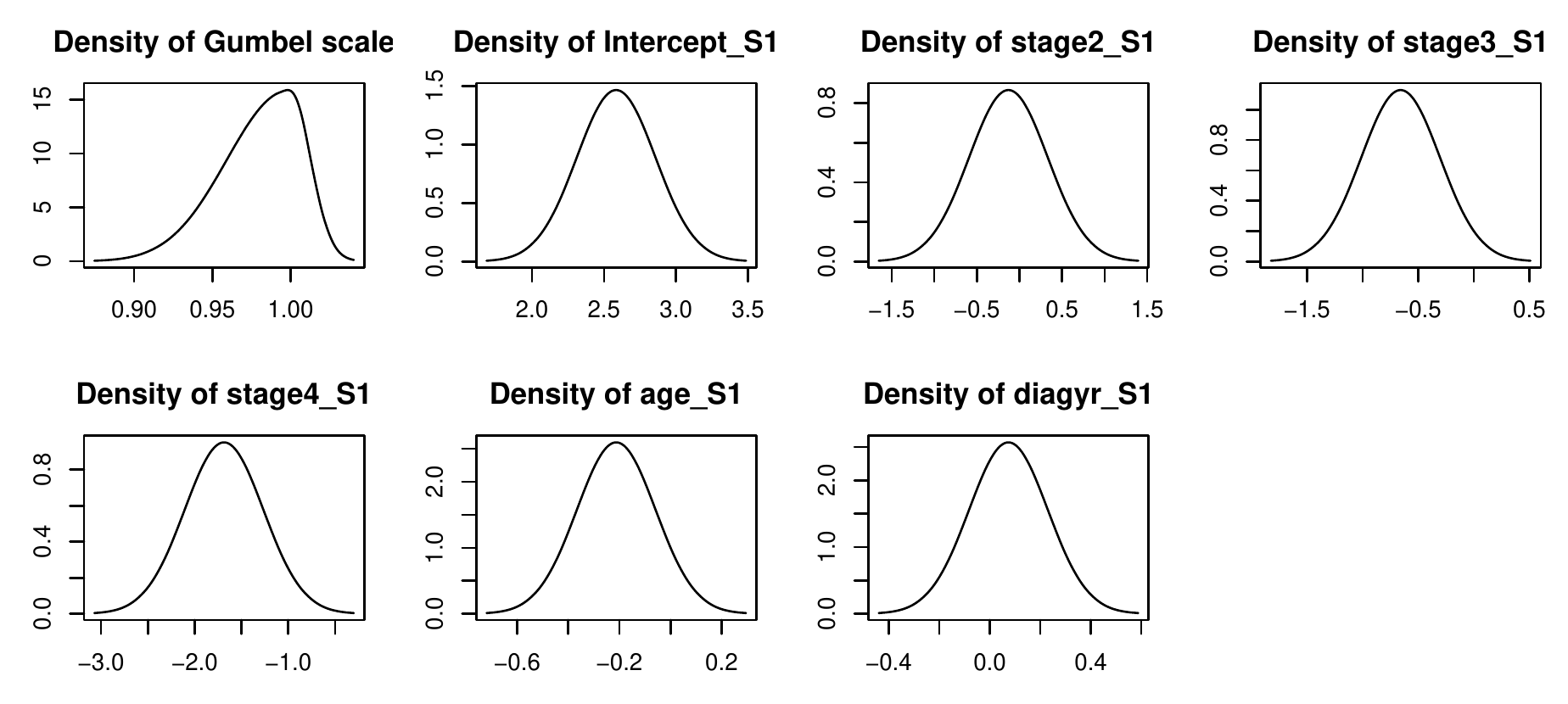}
\caption{Marginal posterior density plots for AFT model parameters.} \label{fig:AFTposterior}
\end{figure}

Figure \ref{fig:AFTposterior} shows the marginal posterior distributions of the model parameters, which can provide additional insight about the distribution of model parameters compared to the summary that focuses on mean, standard deviation, and quantiles (such as visualisation of possible skewness of the distribution). When no transformation is required, one can plot the marginal posterior distributions directly with the \texttt{plot()} function, as illustrated for a joint model in Section \ref{sec:JM1}.

Posterior samples of the hyperparameters and the latent field can be generated using the following command:
\begin{verbatim}
R> m1.sample <- inla.posterior.sample(n, m1.weib)
\end{verbatim}

\noindent where $n$ indicates the number of samples required. Note that the model needs to save some additional information in order to be able to sample from the posterior, therefore the option ``\texttt{config = TRUE}'' must be added to control options of the \texttt{joint()} function when fitting the model.

\subsection{Model interpretation} \label{subsec:aftint}

In the AFT model, the fixed effects $\beta$ represent the log-linear effect of covariates on the survival time. When positive, it suggests that an increase in a given covariate value is associated with a decreasing risk of event (i.e., increases the time to failure) while a negative $\beta$ value suggests that an increase in a given covariate decreases the time to failure. It is although usually preferred to interpret the multiplicative effect of $\exp(\beta)$ on survival. Because we apply an exponential transformation, which is non-linear, it is important to consider the mean of the exponentiated marginal posterior distribution instead of just exponentiating the mean posterior value. For example, for the parameter $\beta_4$ associated with stage 4 of the disease:
\begin{verbatim}
R> inla.zmarginal(inla.tmarginal(function(x) exp(-x), m1.weib$marginals.fixed[["stage4_S1"]]))
      mean         sd quant0.025  quant0.25   quant0.5  quant0.75 quant0.975 
    0.2020     0.0878     0.0813     0.1393     0.1850     0.2457     0.4208 
\end{verbatim}

Having cancer at stage 4 accelerates the time to event by a factor 0.20 [0.08, 0.43] compared to the reference stage 1 (i.e., 0.20 times shorter survival time compared to the baseline survival). Note how this is different from simply exponentiating $\beta$ as $\exp(\hat\beta_4)=\exp(-1.6866)=0.1851$. The location parameter $\exp(\beta_1)$ represents the center of the Gumbel distribution of the survival time for the reference individual. A positive value is associated with longer survival times while a negative value results in shorter survival times. Finally, the interpretation of the Gumbel scale parameter is related to the spread of the baseline survival time distribution, when greater than 1, the survival times are more dispersed. It suggests that there is heterogeneity in the data with extreme or long-surviving cases. On the other hand, the survival times are more concentrated when this parameter is less than 1, meaning that survival times are more similar or clustered around a central value. 
\section{Proportional hazards models} \label{sec:PH}

\subsection{Model specification}
\label{phmodspec}
We will again use the {\it larynx} dataset (see Section~\ref{subsec:aftdata}). In a proportional hazard (PH) approach, the term to be modelled is the hazard function at time $t$:
\begin{align}
h(t \mid {\bm \lambda}, {\bm \beta}) &= h_{0}(t \mid {\bm \lambda}) \exp\big\{ \beta_{2}\mbox{1}_{(\texttt{stage}=2)} + \beta_{3}\mbox{1}_{(\texttt{stage}=3)} + \beta_{4}\mbox{1}_{(\texttt{stage}=4)} + \beta_{5}\texttt{age} +\beta_{6}\texttt{diagyr}\big\}, \quad t > 0, \label{eq:larynx2}
\end{align}

\noindent where $\mbox{1}_{(\texttt{stage}=\cdot)}$ is an indicator variable for \texttt{stage}=2, 3, 4 with regression coefficients $\beta_{2}$, $\beta_{3}$, and $\beta_{4}$, respectively (\texttt{stage}=1 is considered as the reference category); and $\beta_{5}$ and $\beta_{6}$ are regression coefficients for \texttt{age} and \texttt{diagyr} covariates, respectively. Here, the baseline hazard function is set as a mixture of piecewise constant functions, $h_{0}(t \mid {\bm \lambda})= \sum_{k=1}^K\, \lambda_{k} \,\mbox{I}_{(a_{k-1},a_{k}]}(t)$, where $\mbox{I}_{(a_{k-1},a_{k}]}(t)$ is the indicator function defined as $1$ when $t \in (a_{k-1},a_{k}]$ and $0$ otherwise. In this application, we use $K=3$ knots and an equally-spaced partition of the time axis, $a_{0}=0$, $a_{1}=3.57$, $a_{2}=7.13$, and $a_{3}=10.70$. Prior distributions are set according to Section~\ref{subsec:inlajointpriors}.

\subsection{Model implementation}

After loading and transforming the data as in Section~\ref{subsec:aftimpl}, we fit the PH model with an INLA survival object outcome:
\begin{verbatim}
R> m2.pwc <- joint(formSurv = inla.surv(time = time, event = delta) ~ stage + age + diagyr,
+                  basRisk = "rw1", NbasRisk = 3, dataSurv = larynx)
\end{verbatim}

Note that the baseline hazard function (``basRisk'') is specified as a random walk of order one prior (``rw1'')\cite{martino2011}, which is based on first-order differences, i.e., a mixture of piecewise constant functions, where ``NbasRisk'' represents the number of knots. The posterior distribution summary for regression coefficients and baseline risk function variance is given by:
\begin{verbatim}
R> summary(m2.pwc)
Survival outcome
                            mean     sd 0.025quant 0.5quant 0.975quant
Baseline risk (variance)  0.0228 0.0539     0.0001   0.0058     0.1576
stage2                    0.1269 0.4613    -0.7777   0.1269     1.0314
stage3                    0.6530 0.3535    -0.0401   0.6530     1.3461
stage4                    1.6954 0.4222     0.8677   1.6953     2.5236
age                       0.2133 0.1541    -0.0888   0.2133     0.5154
diagyr                   -0.0660 0.1571    -0.3740  -0.0661     0.2423

log marginal-likelihood (integration)    log marginal-likelihood (Gaussian) 
                            -161.5613                             -161.5613 

Deviance Information Criterion:  297.5805
Widely applicable Bayesian information criterion:  299.4084
Computation time: 0.66 seconds
\end{verbatim}

One can also get the posterior distribution summary for $\lambda$'s:
\begin{verbatim}
R> summary(m2.pwc)$BaselineValues
        time  lower median  upper
[1,]  0.0000 0.0436 0.0750 0.1286
[2,]  3.5667 0.0462 0.0809 0.1434
[3,]  7.1333 0.0422 0.0799 0.1560
[4,] 10.7000 0.0383 0.0798 0.1747
\end{verbatim}

\subsection{Model interpretation} \label{subsec:phint}

While in an AFT model the covariates act multiplicatively on time, in a PH model the covariates act multiplicatively on the hazard. Therefore, $\beta$ parameters represent the log hazard ratio associated with a one-unit change in the associated covariate. As for the AFT model, it is more common to exponentiate the $\beta$ parameters to obtain hazard ratios. This can easily be done with \texttt{INLAjoint} by adding the argument ``\texttt{hr = TRUE}'' to the call of the summary or plot function, so the summary statistics are given on the exponential scale (note that baseline parameters are not transformed, as it is not meaningfu
l):
\begin{verbatim}
R> summary(m2.pwc, hr = TRUE)
Survival outcome
                         exp(mean)     sd 0.025quant 0.5quant 0.975quant
Baseline risk (variance)    0.0228 0.0539     0.0001   0.0058     0.1576
stage2                      1.2582 0.5946     0.4638   1.1342     2.7708
stage3                      2.0410 0.7272     0.9681   1.9201     3.8065
stage4                      5.9393 2.5537     2.4024   5.4439    12.3344
age                         1.2521 0.1908     0.9183   1.2374     1.6674
diagyr                      0.9474 0.1473     0.6905   0.9358     1.2688
\end{verbatim}
 
An individual with stage 4 cancer has a hazard 5.94 [2.40, 12.34] times higher than an individual with stage 1, meaning they are nearly six times more likely to die, at any given point in time. The baseline risk represents the risk for the reference individual, here it is assumed piecewise constant with 3 intervals (to match the example previously proposed in Alvares et al\cite{alvares2021}) but the default number of intervals in \texttt{INLAjoint} is 15. Having more intervals leads to more flexibility but can overfit the data. An alternative approach consists in using random walks of order 2, which can be seen as Bayesian smooth splines, to have flexible curves while avoiding parametric assumptions and overfitting, see Section \ref{sec:JM2} for more details.

\subsection{Stratified proportional hazards model} \label{subsec:phstr}
In some situations, the proportional hazards assumption may not hold for some covariates, meaning that the hazard ratio may not be constant in time. In this context, it is possible to stratify the model so that instead of having a proportional effect of a covariate on the risk, we can have a distinct baseline hazard function for each group defined by this covariate. To illustrate this feature in INLAjoint, let's assume the hazards are not proportional conditional on the stage of cancer in the model described in Section \ref{phmodspec}. The model is then defined as:
$$h(t \mid {\bm \lambda}, {\bm \beta}) = h_{0}^{stage_k}(t \mid {\bm \lambda}) \exp\big\{\beta_{1}\texttt{age} +\beta_{2}\texttt{diagyr}\big\}, \quad t > 0,$$
where $h_{0}^{stage_k}(t \mid {\bm \lambda})$ corresponds to the baseline hazard function for stage $k$. The corresponding R code to fit this model then simply includes the covariate ``stage'' as a strata in control options instead of having it directly in the formula:
\begin{verbatim}
R> m2b.pwc <- joint(formSurv = inla.surv(time = time, event = delta) ~ age + diagyr,
+                   dataSurv = larynx, control=list(strata=list("stage")))
\end{verbatim}

The plot function allows to display the baseline hazard function for each strata:
\begin{verbatim}
R> plot(m2b.pwc)
\end{verbatim}
\begin{figure}[ht]
\centering
\includegraphics[scale=0.75]{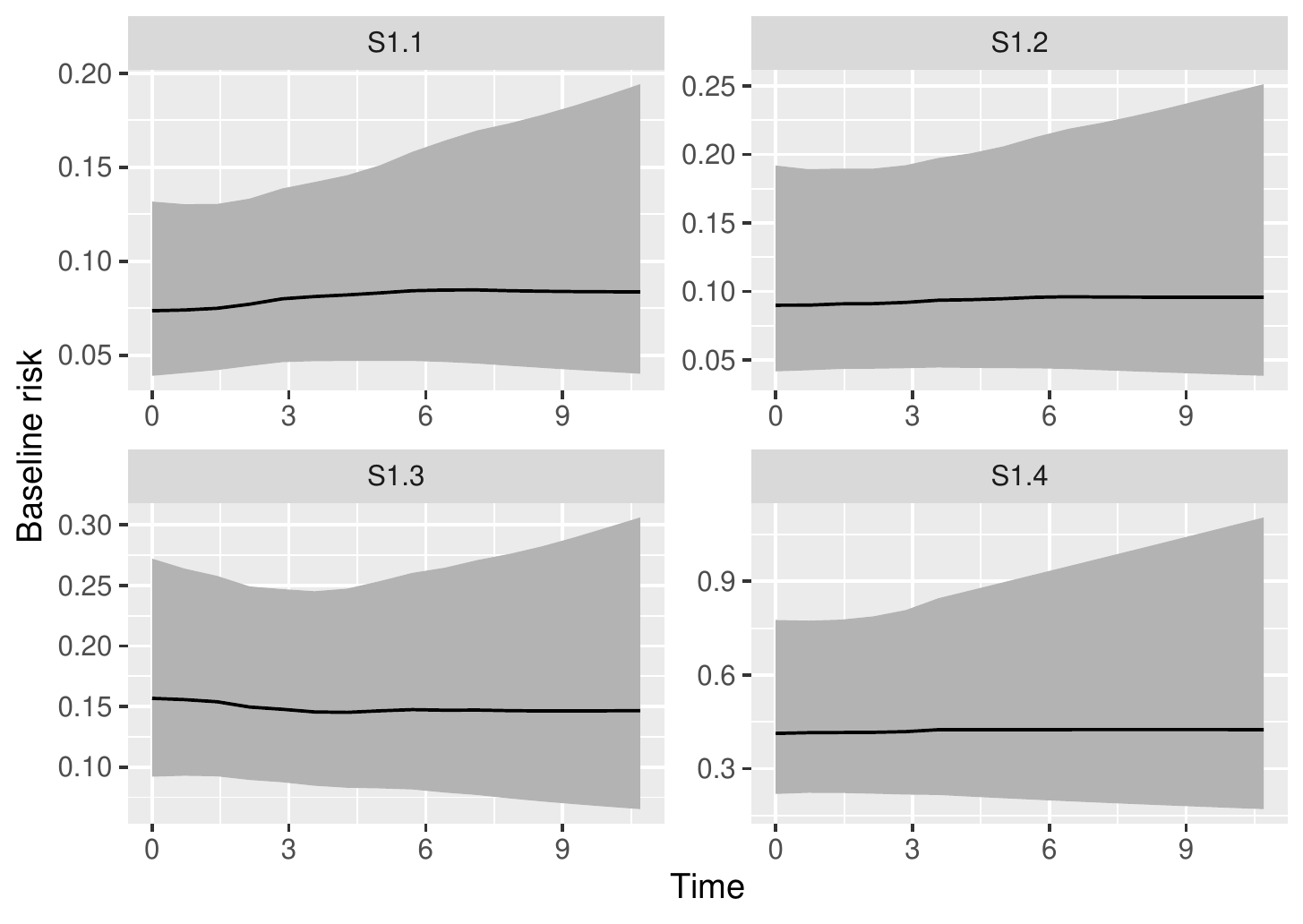}
\caption{Baseline hazard function of each of the 4 stages of cancer for the stratified proportional hazards model.} \label{fig:stratCox}
\end{figure}

Figure \ref{fig:stratCox} shows the stata-specific baseline hazard functions. Here, we can see that although the Y-axis scale is different for each baseline hazard function (i.e., higher stage of cancer corresponds to more advanced disease and thus higher risk of death), there is no evidence for non-proportional hazards conditional on the stage covariate and the results are in line with results from the non-stratified proportional hazards model. Allowing for non-proportional hazards conditional on the cancer stage does not seem to provide benefit here but when baseline hazard curves are not parallel, it would indicate non-proportionality conditional on the strata covariate.

\section{Mixture cure models} \label{sec:Cure}

\subsection{Motivation: bmt dataset (from the \texttt{smcure} R-package)}

The {\it bmt} dataset contains 91 patients with refractory acute lymphoblastic leukemia who participated in a bone marrow transplant study.\cite{kersey1987} Each patient has the following variables:
\begin{itemize}
	\item \texttt{Time}: time to death (in days).
	\item \texttt{Status}: censoring indicator (1: if patient is uncensored; 0: otherwise).
	\item \texttt{TRT}: treatment group indicator (1: autologous; 0: allogeneic).
\end{itemize}

Of the 91 patients, 46 received allogeneic transplant (treatment); and the observed survival times had 11 days as minimum, 1845 as maximum, and 179 as median, with 24.2\% right-censored.

\subsection{Model specification}

Let $Z$ be a binary random variable defined as $0$ for susceptible (i.e., $Z=1-\texttt{Status}$) and $1$ for cured or immune individuals. So, the incidence submodel is given by:
\begin{align}
 Z &\sim \mbox{Bernoulli}(\eta), \label{eq:bmt11} \\
 \mbox{logit}(\eta) &= \beta_{C1} + \beta_{C2}\texttt{TRT}, \label{eq:bmt12}
\end{align}

\noindent where $\beta_{C1}$ represents an intercept and $\beta_{C2}$ is the regression coefficient for the \texttt{TRT} covariate.

The latency submodel is expressed through a PH specification:
\begin{equation}
h(t \mid \lambda, \alpha, \beta_{U}) = h_{0}(t \mid \lambda, \alpha)\exp\left\{\beta_{U}\texttt{TRT}\right\}, \quad t > 0, \label{eq:bmt2}
\end{equation}

\noindent with $h_{0}(t \mid \lambda, \alpha)=\lambda \, \alpha \, t^{\alpha-1}$ specified as a Weibull baseline hazard function, where $\lambda$ and $\alpha$ are the scale and shape parameters, respectively; and $\beta_{U}$ is the regression coefficient for the \texttt{TRT} covariate. Prior distributions are set according to Section~\ref{subsec:inlajointpriors}.

\subsection{Model implementation}

We start by loading the data:
\begin{verbatim}
R> library(smcure)
R> data(bmt)
\end{verbatim}

So, we setup an INLA survival object indicating that the incidence submodel (``cure'') contains an intercept and the \texttt{TRT} covariate and fit the mixture cure model:
\begin{verbatim}            
R> m3.cure <- joint(formSurv = inla.surv(time = Time, event = Status,
+                                        cure = cbind("Int"=1, "TRT"=TRT)) ~ TRT,
+                   basRisk = "weibullsurv", dataSurv = bmt)
\end{verbatim}

The posterior distribution summary is given by:
\begin{verbatim}
R> summary(m3.cure)
Survival outcome
                   mean     sd 0.025quant 0.5quant 0.975quant
Int(cure)       -1.0019 0.3220    -1.6355  -1.0020    -0.3677
TRT(cure)       -0.3966 0.4555    -1.2996  -0.3945     0.4938
Weibull (shape)  1.0921 0.0883     0.9414   1.0847     1.2877
Weibull (scale)  0.0027 0.0009     0.0013   0.0026     0.0046
TRT              0.7028 0.2556     0.2014   0.7030     1.2038

log marginal-likelihood (integration)    log marginal-likelihood (Gaussian) 
                            -487.1486                             -487.1486 

Deviance Information Criterion:  954.4586
Widely applicable Bayesian information criterion:  955.6154
Computation time: 0.67 seconds
\end{verbatim}

\subsection{Model interpretation} \label{subsec:Cureint}

In order to interpret the model, we first need to compute the cure fraction of the model, with \texttt{INLAjoint} the parameters of the incidence submodel are considered as hyperparameters and we can easily sample them with the function \texttt{inla.hyperpar.sample}:
\begin{verbatim}
R> smp.cure <- inla.hyperpar.sample(500, m3.cure)[, 2:3]
R> quantile(inv.logit(smp.cure[,1]), c(0.025, 0.5, 0.975)) # allogeneic
  2.5%    50%  97.5% 
0.1595 0.2641 0.4075
R> quantile(inv.logit(rowSums(smp.cure)), c(0.025, 0.5, 0.975)) # autologous
  2.5%    50%  97.5% 
0.1045 0.1970 0.3272
\end{verbatim}

Therefore the proportion of the population that is ``cured'' and will not experience the event is 26\% [15\%, 41\%] for the treatment group allogeneic and 20\% [10\%, 33\%] for the treatment group autologous. For the remaining patients who are not considered cured, we can estimate the hazard ratio of treatment:
\begin{verbatim}
R> summary(m3.cure, hr=T)$SurvEff[[1]]["TRT_S1",]
       exp(mean)     sd 0.025quant 0.5quant 0.975quant
TRT_S1     2.084 0.5310     1.2302   2.0189     3.3099
\end{verbatim}

In the non-cured fraction of patients, those who received autologous treatment have a hazard of experiencing the event 2.08 [1.23, 3.31] times higher than those who received allogeneic treatment, indicating a significantly greater risk associated with the autologous treatment option.
\section{Competing risks models} \label{sec:CR}

\subsection{Motivation: okiss dataset (from the \texttt{compeir} R-package)}

The {\it okiss} dataset contains a random subsample of 1000 patients from ONKO-KISS, which is part of the surveillance program of the German National Reference Centre for Surveillance of Hospital-Acquired Infections.\cite{dettenkofer2005} After a peripheral blood stem-cell transplantation, these patients are neutropenic (low number of white blood cells). Each patient has the following variables:
\begin{itemize}
	\item \texttt{time}: time (in days) of neutropenia until first event.
	\item \texttt{status}: event status indicator (1: infection; 2: end of neutropenia; 7: death; 11: censored observation).
	\item \texttt{allo}: transplant type indicator (1: allogeneic; 0: autologous).
	\item \texttt{sex}: sex of each patient (m: if patient is male; f: if patient is female).
\end{itemize}

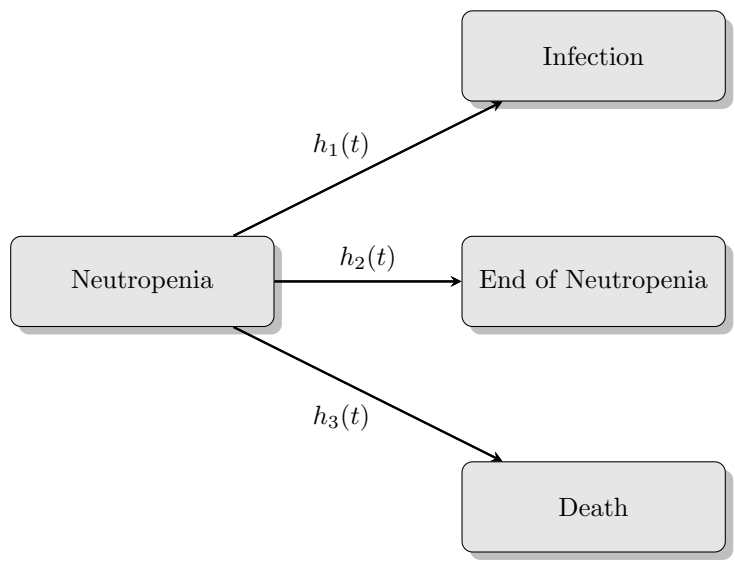
\begin{figure}[ht]
\centering
\begin{tikzpicture}[node distance=3cm]
\node [drop shadow={shadow xshift=0.3em, shadow yshift=-0.3em}] (state1) [startstop] {Neutropenia};
\node [drop shadow={shadow xshift=0.3em, shadow yshift=-0.3em}] (state3) [startstop, right of=state1, xshift=3cm] {End of Neutropenia};
\node [drop shadow={shadow xshift=0.3em, shadow yshift=-0.3em}] (state2) [startstop, above of=state3] {Infection};
\node [drop shadow={shadow xshift=0.3em, shadow yshift=-0.3em}] (state4) [startstop, below of=state3] {Death};
\draw [arrow] (state1) -- (state2);
\draw [arrow] (state1) -- (state3);
\draw [arrow] (state1) -- (state4);
\draw [arrow] (state1) -- node[anchor=south] {$h_{1}(t) \qquad$} (state2);
\draw [arrow] (state1) -- node[anchor=south] {$h_{2}(t)$} (state3);
\draw [arrow] (state1) -- node[anchor=north] {$h_{3}(t) \qquad$} (state4);
\end{tikzpicture}
\caption{Graphical representation of competing risks for {\it okiss} dataset.  All the patients enter the follow-up with neutropenia and are at risk to have an infection, the end of neutropenia or death if they are not censored before observing one of these three competing events.} \label{fig:CR}
\end{figure}

Of the 1000 patients, 564 received allogeneic transplant; 61.9\% were male; and the median survival times per event were 6 (infection), 12 (end of neutropenia), and 15 (death), with 13\% right-censored. Figure~\ref{fig:CR} illustrates the competitive risks transition structure for {\it okiss} dataset.

\subsection{Model specification}

Cause-specific hazard functions for infection ($k=1$), end of neutropenia ($k=2$), and death ($k=3$) are modelled from a PH specification:
\begin{align}
	h_{1}(t \mid \lambda_{1}, \alpha_{2}, {\bm\beta}_{1}) &= h_{01}(t \mid \lambda_{1}, \alpha_{2})\exp\left\{\beta_{11}\texttt{allo} + \beta_{21}\texttt{sex}\right\}, \quad t > 0, 		  \label{eq:cr1} \\
	h_{2}(t \mid \lambda_{2}, \alpha_{2}, {\bm\beta}_{2}) &= h_{02}(t \mid \lambda_{2}, \alpha_{2})\exp\left\{\beta_{12}\texttt{allo} + \beta_{22}\texttt{sex}\right\}, \quad t > 0, 		  \label{eq:cr2} \\
	h_{3}(t \mid \lambda_{3}, \alpha_{3}, {\bm\beta}_{3}) &= h_{03}(t \mid \lambda_{3}, \alpha_{3})\exp\left\{\beta_{13}\texttt{allo} + \beta_{23}\texttt{sex}\right\}, \quad t > 0, \label{eq:cr3}
\end{align}

\noindent with $h_{0k}(t \mid \lambda_{k}, \alpha_{k})=\lambda_{k} \, \alpha_{k} \, t^{\alpha_{k}-1}$ for event $k$ specified as a Weibull baseline hazard function, where $\lambda_{k}$ and $\alpha_{k}$ are the scale and shape parameters, respectively; and ${\bm \beta}_{k}=(\beta_{1k},\beta_{2k})^{\top}$ are regression coefficients for the \texttt{allo} and \texttt{sex} covariates, respectively, for $k=1,2,3$. Prior distributions are set according to Section~\ref{subsec:inlajointpriors}.

\subsection{Model implementation}

We start by loading the data and creating an auxiliary variable (\texttt{delta}) from \texttt{status}:
\begin{verbatim}
R> library(compeir)
R> data(okiss)
R> delta <- matrix(c(as.integer(okiss$status == 1),
+                    as.integer(okiss$status == 2),
+                    as.integer(okiss$status == 7)), ncol = 3)
R> head(delta)
     [,1] [,2] [,3]
[1,]    0    1    0
[2,]    1    0    0
[3,]    0    1    0
[4,]    0    1    0
[5,]    0    1    0
[6,]    0    1    0
\end{verbatim}

In this auxiliary variable, the events 1 (infection), 2 (end of neutropenia), or 7 (death) are indicated with a value of 1 in columns 1, 2, or 3, respectively, and a row with only 0's represents a censored observation.

The next step is to setup an INLA survival object for each hazard function and fit the competing risks model:
\begin{verbatim}
R> m4.cr <- joint(formSurv = list(inla.surv(time = time, event = delta[,1]) ~ allo + sex, 
+                                 inla.surv(time = time, event = delta[,2]) ~ allo + sex, 
+                                 inla.surv(time = time, event = delta[,3]) ~ allo + sex), 
+                 basRisk = rep("weibullsurv", 3), dataSurv = okiss)
\end{verbatim}

Note that the regressors and the baseline hazard of each hazard function are set independently. Finally, the posterior distribution summary is given by:
\begin{verbatim}
R> summary(m4.cr)
Survival outcome (S1)
                      mean     sd 0.025quant 0.5quant 0.975quant
Weibull (shape)_S1  1.1030 0.0638     0.9829   1.1011     1.2339
Weibull (scale)_S1  0.0156 0.0026     0.0111   0.0154     0.0213
allo_S1            -0.5032 0.1443    -0.7861  -0.5032    -0.2202
sexm_S1             0.1547 0.1477    -0.1349   0.1547     0.4443

Survival outcome (S2)
                      mean     sd 0.025quant 0.5quant 0.975quant
Weibull (shape)_S2  2.0278 0.0445     1.9414   2.0273     2.1166
Weibull (scale)_S2  0.0077 0.0007     0.0064   0.0077     0.0092
allo_S2            -1.1919 0.0734    -1.3357  -1.1919    -1.0480
sexm_S2            -0.1009 0.0737    -0.2453  -0.1009     0.0436

Survival outcome (S3)
                      mean     sd 0.025quant 0.5quant 0.975quant
Weibull (shape)_S3  2.2053 0.2848     1.6759   2.1946     2.7939
Weibull (scale)_S3  0.0000 0.0000     0.0000   0.0000     0.0001
allo_S3            -0.4756 0.6729    -1.7950  -0.4756     0.8440
sexm_S3             0.4677 0.6672    -0.8407   0.4677     1.7760

log marginal-likelihood (integration)    log marginal-likelihood (Gaussian) 
                            -3735.786                             -3735.786 

Deviance Information Criterion:  7409.673
Widely applicable Bayesian information criterion:  7421.033
Computation time: 1.1 seconds
\end{verbatim}

\subsection{Model interpretation} \label{subsec:CRint}

For each submodel, the Weibull distribution describes how the hazard of the event of interest changes with time. For example the Weibull shape parameter for the death event, $\hat\alpha_3$ = 2.21 [1.67, 2.80], suggests that the hazard of death increases over time while the hazard of infection is more stable ($\hat\alpha_1$ = 1.10 [0.98, 1.24]). The effect of covariates is straightforward, for example, the hazard ratio of allogeneic versus autologous transplant type for the risk of infection is obtained as follows:
\begin{verbatim}
R> summary(m4.cr, hr=T)$SurvEff[[1]]["allo_S1",]
        exp(mean)     sd 0.025quant 0.5quant 0.975quant
allo_S1    0.6107 0.0871     0.4571   0.6045     0.7992
\end{verbatim}

This suggests that patients receiving allogeneic transplants have a 39\% [20\%, 55\%] lower risk of infection compared to patients receiving autologous transplants, accounting for other competing risks. We can also plot the {\it cumulative incidence function} (CIF), which describes the cumulative probability of experiencing a specific event of interest over time while accounting for the presence of other competing events. For example for the reference individual (i.e., female with autologous transplant type):
\begin{verbatim}
R> riskW <- function(t, lambda, alpha) lambda*alpha*t^(alpha-1)
R> t <- seq(0, 100, len=100)
R> risk1 <- riskW(t, exp(m4.cr$summary.fixed["Intercept_S1", "mean"]), 
R>                m4.cr$summary.hyperpar$mean[1])
R> risk2 <- riskW(t, exp(m4.cr$summary.fixed["Intercept_S2", "mean"]), 
R>                m4.cr$summary.hyperpar$mean[2])
R> risk3 <- riskW(t, exp(m4.cr$summary.fixed["Intercept_S3", "mean"]), 
R>                m4.cr$summary.hyperpar$mean[3])
R> surv <- exp(-cumsum(risk1) - cumsum(risk2) - cumsum(risk3))
R> CIF <- cbind(cumsum(risk1*surv), cumsum(risk2*surv), cumsum(risk3*surv))
\end{verbatim}

\begin{figure}[ht]
\centering
\includegraphics[scale=0.75]{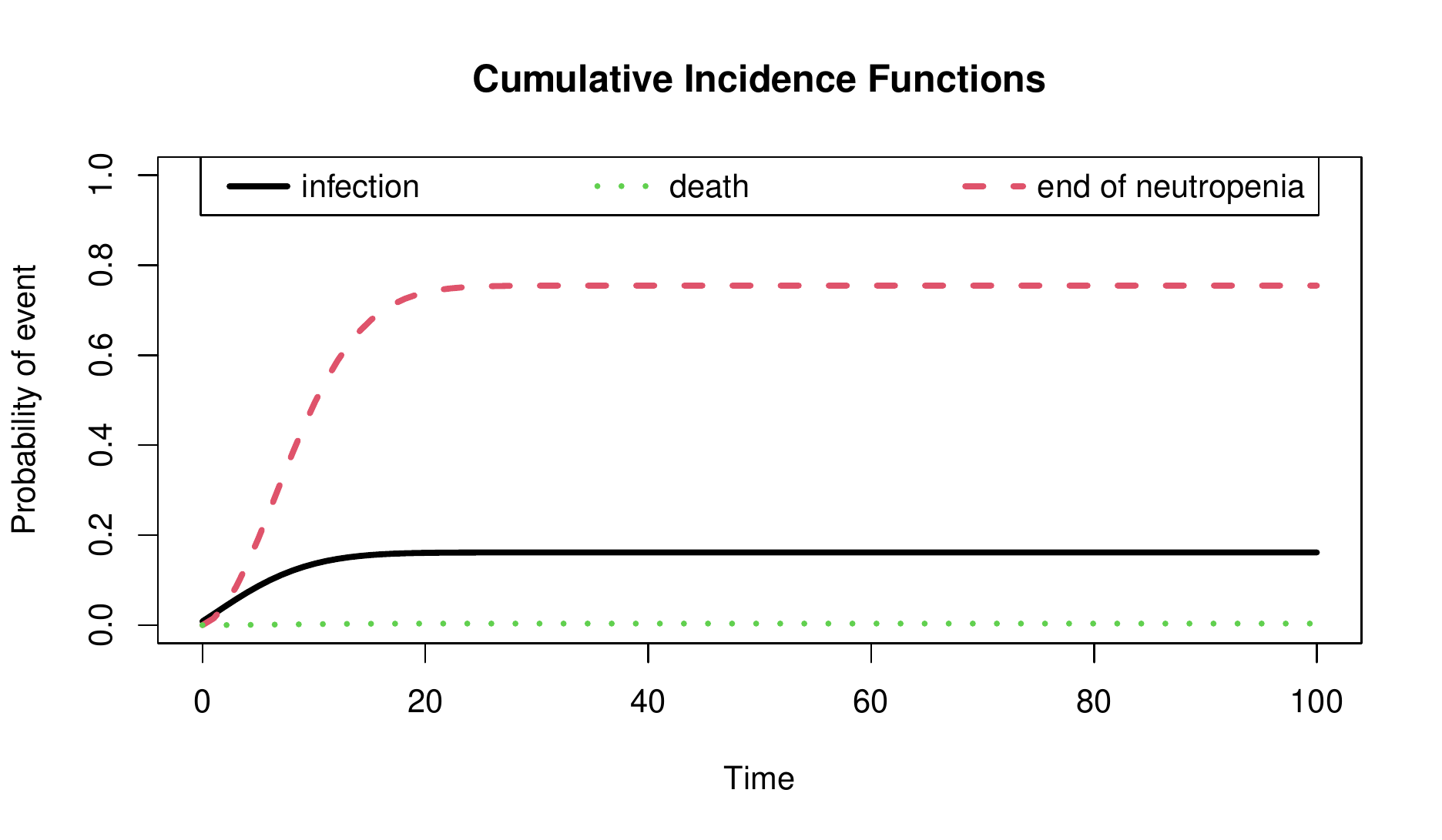}
\caption{Cumulative incidence functions of the reference individual (i.e., female with autologous transplant type) for the competing risks model.}
\label{CIF}
\end{figure}

CIFs are displayed in Figure \ref{CIF}, showing a higher probability of experiencing the end of neutropenia compared to other competing events over time. Note that it is possible to add uncertainty by sampling parameters as described in Section \ref{subsec:aftimpl}. Note that CIFs are preferred over survival curves in the context of competing risks as survival curves would represent the probability of having an event in a hypothetical world where it is not possible to have any of the competing events.

\section{Multi-state models} \label{sec:MS}

\subsection{Motivation: heart2 dataset (from the \texttt{p3state.msm} R-package)}

The {\it heart2} dataset contains a sample of 103 patients of the Stanford Heart Transplant Program.\cite{crowley1977} The patients are on the waiting list (initial state) and can either be transplanted (non-terminal event) and then die (terminal event), or just one or none of them because they continue to be on the waiting list. Each patient has the following variables:
\begin{itemize}
	\item \texttt{times1}: time of transplant/censoring time.
	\item \texttt{delta}: transplant indicator (1: yes; 0: no).
	\item \texttt{times2}: time to death since the transplant/censoring time.
	\item \texttt{time}: \texttt{times1} + \texttt{times2}.
	\item \texttt{status}: censoring indicator (1: dead; 0: alive).
	\item \texttt{age}: age - 48 years.
	\item \texttt{year}: year of acceptance (in years after 1 Nov 1967).
	\item \texttt{surgery}: prior bypass surgery (1: yes; 0: no).		
\end{itemize}

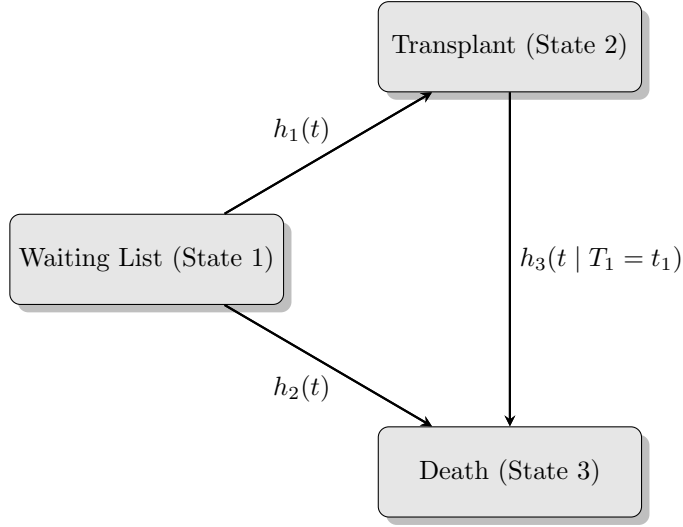
\begin{figure}[ht]
\centering
\begin{tikzpicture}[node distance=4cm]
\node [drop shadow={shadow xshift=0.3em, shadow yshift=-0.3em}] (state1) [startstop] {Waiting List (State 1)};
\node [drop shadow={shadow xshift=0.3em, shadow yshift=-0.3em}] (state2) [startstop, above right of=state1, xshift=2cm] {Transplant (State 2)};
\node [drop shadow={shadow xshift=0.3em, shadow yshift=-0.3em}] (state3) [startstop, below right of=state1, xshift=2cm] {Death (State 3)};
\draw [arrow] (state1) -- (state2);
\draw [arrow] (state1) -- (state3);
\draw [arrow] (state2) -- (state3);
\draw [arrow] (state1) -- node[anchor=south] {$h_{1}(t) \qquad$} (state2);
\draw [arrow] (state1) -- node[anchor=north] {$h_{2}(t) \qquad$} (state3);
\draw [arrow] (state2) -- node[anchor=west] {$h_{3}(t \mid T_{1}=t_{1}) \qquad$} (state3);
\end{tikzpicture}
\caption{Graphical representation of (illness-death) multi-state for {\it heart2} dataset. Patients enter the follow-up while being on the waiting list for transplantation. They can either receive transplantation and die after their transplantation or directly die while still being on the waiting list. Additionally, censoring can occur at anytime during the follow-up.}\label{fig:MS}
\end{figure}

The patients' ages (minus 48) were distributed between -39.2 and 16.4 with a mean of -2.8 and a standard deviation of 9.8; the year of acceptance ranged approximately uniformly from -0.05 to 6.47; and 84.5\% of patients had not undergone prior bypass surgeries. In addition, 30 patients died before transplantation, 69 were transplanted, of which 45 died after transplantation, and only 4 patients did not experience either event (censored). Figure~\ref{fig:MS} illustrates the multi-state transition structure for {\it heart2} dataset.

\subsection{Model specification}

Hazard functions from waiting list to transplant ($h_{1}$), from waiting list to death ($h_{2}$), and from waiting list to death after transplant ($h_{3}$) are modelled from a PH specification:
\begin{align}
	h_{1}(t \mid \lambda_{1}, \alpha_{1}, {\bm\beta}_{1}) &= h_{01}(t \mid \lambda_{1}, \alpha_{1})\exp\left\{\beta_{12}\texttt{age}+\beta_{13}\texttt{year}+\beta_{14}\texttt{surgery}\right\}, \quad t > 0,  \label{eq:scr1} \\
	h_{2}(t \mid \lambda_{2}, \alpha_{2}, {\bm\beta}_{2}) &= h_{02}(t \mid \lambda_{2}, \alpha_{2})\exp\left\{\beta_{22}\texttt{age}+\beta_{23}\texttt{year}+\beta_{24}\texttt{surgery}\right\}, \quad t > 0,  \label{eq:scr2} \\
	h_{3}(t \mid \lambda_{3}, \alpha_{3}, {\bm\beta}_{3}, T_{1}=t_{1}) &= h_{03}(t-t_{1} \mid \lambda_{3}, \alpha_{3})\exp\left\{\beta_{32}\texttt{age}+\beta_{33}\texttt{year}+\beta_{34}\texttt{surgery}\right\}, \quad t > t_{1},  \label{eq:scr3}
\end{align} \normalsize

\noindent with $h_{0k}(t \mid \lambda_{k}, \alpha_{k})=\lambda_{k} \, \alpha_{k} \, t^{\alpha_{k}-1}$ specified as a Weibull baseline hazard function, where $\lambda_{k}$ and $\alpha_{k}$ are the scale and shape parameters, respectively; and ${\bm\beta}_{k}=(\beta_{1k},\beta_{2k},\beta_{3k})^{\top}$ are regression coefficients for the \texttt{age}, \texttt{year} and \texttt{surgery} covariates, respectively, for $k=1,2,3$. Note that we adopted a semi-Markovian specification in \eqref{eq:scr3}.\cite{alvares2019} In this case, the model can also be seen as a clock-reset specification,\cite{kleinbaum2012} in which time starts at zero again after each transition. Prior distributions are set according to Section~\ref{subsec:inlajointpriors}.

\subsection{Model implementation}

We start by loading the data, creating an auxiliary variable (\texttt{event}) from \texttt{status}, and setting the time from transplant to death or censoring:
\begin{verbatim}
R> library(p3state.msm)
R> data(heart2)
R> event <- matrix(c(heart2$delta, heart2$status * (1 - heart2$delta), 
+                    heart2$delta * heart2$status), ncol = 3)
R> head(event)
     [,1] [,2] [,3]
[1,]    0    1    0
[2,]    0    1    0
[3,]    1    0    1
[4,]    1    0    1
[5,]    0    1    0
[6,]    0    1    0
R> heart2$times3 <- ifelse(heart2$times2 == 0, heart2$times1, heart2$times2)
\end{verbatim}

Each column in this auxiliary variable represents the transition of a patient. For example, the first row indicates that the patient has passed from waiting list to death, and the third row indicates that the patient has passed from waiting list to transplant and after from transplant to death. A row with only 0's represents a censored observation from waiting list.

Similar to the competing risks model, the next step is to set an INLA survival object for each hazard function and fit the multi-state model:
\begin{verbatim}
R> m5.ms <- joint(formSurv = list(inla.surv(times1, event[,1]) ~ age + year + surgery, 
+                                 inla.surv(times3, event[,2]) ~ age + year + surgery, 
+                                 inla.surv(times2, event[,3]) ~ age + year + surgery),
+                 basRisk = rep("weibullsurv", 3), dataSurv = heart2)
\end{verbatim}

Finally, the posterior distribution summary is given by:
\begin{verbatim}
R> summary(m5.ms)
Survival outcome (S1)
                      mean     sd 0.025quant 0.5quant 0.975quant
Weibull (shape)_S1  0.8174 0.0592     0.7062   0.8156     0.9393
Weibull (scale)_S1  0.0349 0.0105     0.0186   0.0334     0.0596
age_S1              0.0508 0.0140     0.0235   0.0508     0.0782
year_S1            -0.0001 0.0686    -0.1346  -0.0001     0.1345
surgery_S1          0.2301 0.3141    -0.3857   0.2301     0.8460

Survival outcome (S2)
                      mean     sd 0.025quant 0.5quant 0.975quant
Weibull (shape)_S2  0.4947 0.0399     0.4202   0.4933     0.5774
Weibull (scale)_S2  0.0535 0.0201     0.0243   0.0500     0.1026
age_S2              0.0038 0.0179    -0.0313   0.0038     0.0390
year_S2            -0.2226 0.1144    -0.4469  -0.2226     0.0017
surgery_S2         -0.9450 0.6205    -2.1617  -0.9450     0.2718

Survival outcome (S3)
                      mean     sd 0.025quant 0.5quant 0.975quant
Weibull (shape)_S3  0.6699 0.0630     0.5522   0.6677     0.7998
Weibull (scale)_S3  0.0193 0.0077     0.0083   0.0179     0.0382
age_S3              0.0590 0.0215     0.0170   0.0590     0.1012
year_S3             0.0085 0.0913    -0.1705   0.0085     0.1876
surgery_S3         -1.0341 0.4447    -1.9060  -1.0341    -0.1621

log marginal-likelihood (integration)    log marginal-likelihood (Gaussian) 
                            -925.9973                             -925.9973 

Deviance Information Criterion:  1763.492
Widely applicable Bayesian information criterion:  1769.848
Computation time: 0.87 seconds
\end{verbatim}

Alternatively, one can change the transition specification $2\rightarrow3$ ($h_{3}$) in order to enter the at-risk group after the transition $1\rightarrow2$ (no clock-reset) as follows:
\begin{verbatim}
R> surv.obj.ms3 <- inla.surv(time = heart2$time, truncation = heart2$times1, event = event[,3])
\end{verbatim}

Note that now the INLA survival object for hazard function $h_{3}$ includes the "truncation" argument, which indicates the transition time from waiting list to transplant. 

\subsection{Model interpretation} \label{subsec:MSint}

In the context of multi-state models with time-dependent transition probabilities, it is often preferred to employ graphical representations to illustrate the evolution of these transition probabilities over time. It is possible to compute the transition probabilities from the coefficients of the survival submodels fitted with \texttt{INLAjoint}. For example, we compute the probability of each transition conditional on covariate ``surgery'' as follows (we consider the reference individual and assume a transition to state 2 occurred at a time 0 for the intensities p22 and p23 for simplicity):
\begin{verbatim}
R> t <- seq(0.1, 1000, by = 1)
R> riskW <- function(t, lambda, alpha) lambda*alpha*t^(alpha-1)
R> risk1 <- riskW(t, exp(m5.ms$summary.fixed["Intercept_S1", "mean"]),
R>                m5.ms$summary.hyperpar$mean[1])
R> risk2 <- riskW(t, exp(m5.ms$summary.fixed["Intercept_S2", "mean"]),
R>                m5.ms$summary.hyperpar$mean[2])
R> risk3 <- riskW(t, exp(m5.ms$summary.fixed["Intercept_S3", "mean"]),
R>                m5.ms$summary.hyperpar$mean[3])
R> risk1s <- riskW(t, exp(m5.ms$summary.fixed["Intercept_S1", "mean"] +
R>                        m5.ms$summary.fixed["surgery_S1", "mean"]),
R>                 m5.ms$summary.hyperpar$mean[1])
R> risk2s <- riskW(t, exp(m5.ms$summary.fixed["Intercept_S2", "mean"] +
R>                        m5.ms$summary.fixed["surgery_S2", "mean"]),
R>                 m5.ms$summary.hyperpar$mean[2])
R> risk3s <- riskW(t, exp(m5.ms$summary.fixed["Intercept_S3", "mean"] +
R>                        m5.ms$summary.fixed["surgery_S3", "mean"]), 
R>                 m5.ms$summary.hyperpar$mean[3])
R> p11 <- exp(-cumsum(risk1)-cumsum(risk2))
R> p11s <- exp(-cumsum(risk1s)-cumsum(risk2s))
R> p22 <- exp(-cumsum(risk3))
R> p22s <- exp(-cumsum(risk3s))
R> p12 <- cumsum(p11*risk1*p22)
R> p12s <- cumsum(p11s*risk1s*p22s)
R> p13 <- 1-p11-p12
R> p13s <- 1-p11s-p12s
R> p23 <- 1-p22
R> p23s <- 1-p22s
\end{verbatim}

\begin{figure}[ht]
\centering
\includegraphics[scale=0.75]{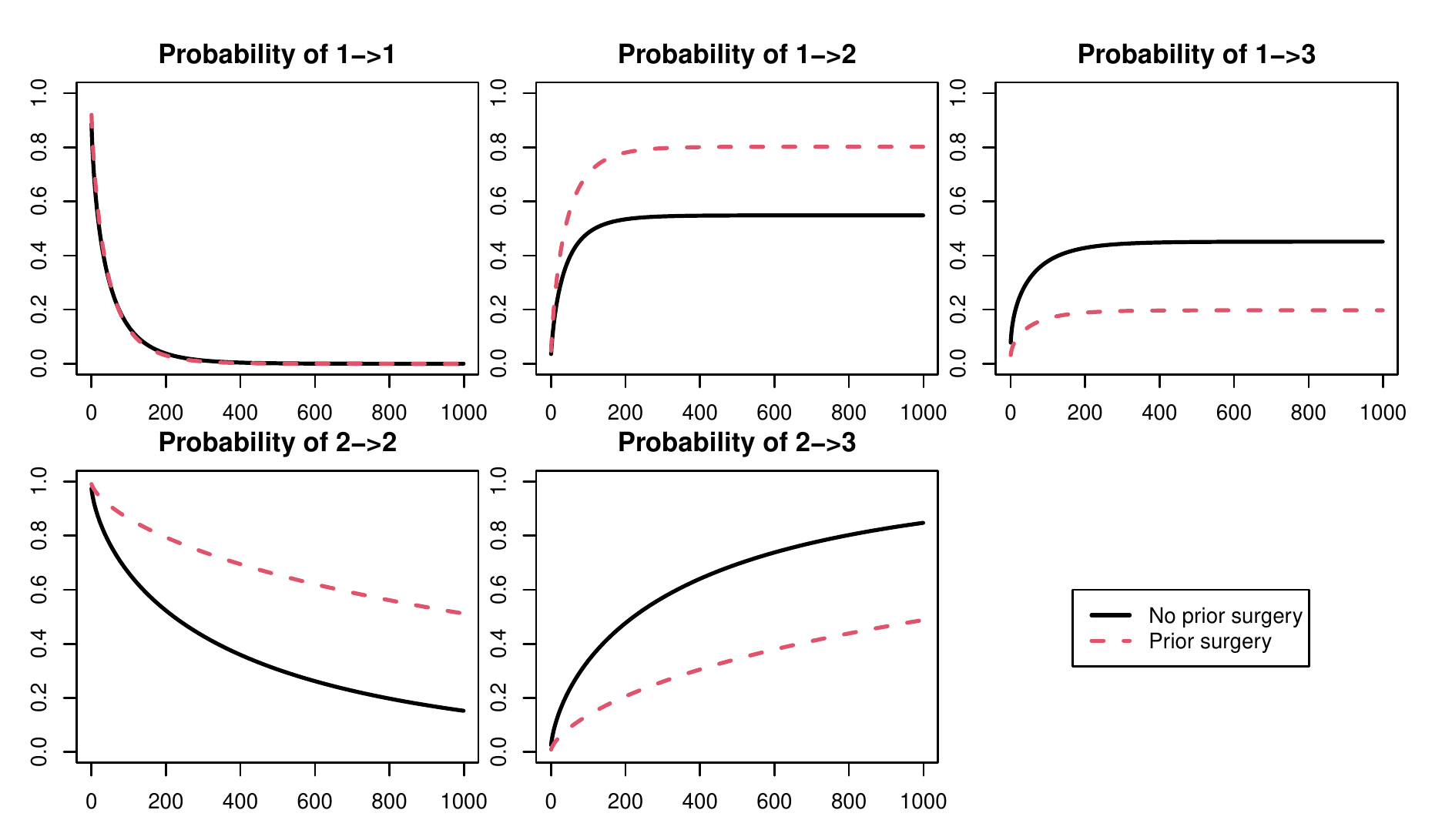}
\caption{Posterior mean of the probability transitions of the multi-state model conditional on prior bypass surgery (assuming the reference individual and transition to state 2 occurred a time 0 for the intensities p22 and p23 for simplicity). The transition probabilities evolve over time but always sum to 1 for a given initial state at a given time point.} Uncertainty over these curves can be obtained through sampling.
\label{Transi}
\end{figure}

We plotted each transition intensity conditional on prior surgery in Figure \ref{Transi}. We can read from this plot that for reference values of age (48 years old) and year of acceptance (1967), patients who received prior bypass surgery are more likely to get transplantation and less likely to experience death before transplantation compared to patients without prior surgery. For patients who did receive transplantation, those who received prior bypass surgery are also less likely to die over time compared to those who did not.

\section{Frailty models} \label{sec:Frailty}

\subsection{Motivation: kidney dataset (from the \texttt{frailtyHL} R-package)}

The \texttt{kidney} dataset contains 38 kidney patients using a portable dialysis machine. \cite{mcgilchrist1991} Times up to the first and second infections are recorded (i.e., each patient has two times-to-infection). Infections can occur at the location of insertion of the catheter. When this is the case, then the catheter is later removed, but it can also be removed for other reasons (censored observation). Each patient has the following variables:
\begin{itemize}
	\item \texttt{id}: patient number.
	\item \texttt{time}: time (in days) from insertion of the catheter to infection in kidney patients using portable dialysis machine.
	\item \texttt{status}: censoring indicator (1: if patient is uncensored; 0: otherwise).
	\item \texttt{age}: age (in years) of each patient.
	\item \texttt{sex}: sex of each patient (1: if patient is male; 2: if patient is female).
	\item \texttt{disease}: disease type (GN; AN; PKD; Other).
	\item \texttt{frail}: frailty estimate from original paper.
	\item \texttt{GN}: indicator for disease type GN.
	\item \texttt{AN}: indicator for disease type AN.
	\item \texttt{PKD}: indicator for disease type PKD.
\end{itemize}

Of the 38 patients, 28 were female; and the median times until the first and second infection were 46 and 39 days, respectively, with 9 right-censored observations. The description of the other variables was omitted because they will not be used in the modelling.

\subsection{Model specification}

Similar to the multi-state model, we will use the clock-reset specification, in which time starts at zero again after the first infection and the term frailty is what connects the two times-to-infection of the same patient. From a PH approach, the hazard function for the frailty model is described as follows:
\begin{equation}
h(t \mid \lambda, \alpha, {\bm \beta}, b_{i}) = h_{0}(t \mid \lambda, \alpha) \exp\left\{\beta\texttt{sex} + b_{i} \right\}, \quad t > 0, \label{eq:kidney}
\end{equation}

\noindent with $h_{0}(t \mid \lambda, \alpha) = \lambda \, \alpha \, t^{\alpha-1}$ specified as a Weibull baseline hazard function, where $\lambda$ and $\alpha$ are the scale and shape parameters, respectively; and $\beta$ is the regression coefficient for the \texttt{sex} covariate. Note that we used an additive frailty specification, where $b_{i}$'s are assumed as normally distributed with zero mean and unknown variance. For simplicity, we do not use the other covariates. Prior distributions are set according to Section~\ref{subsec:inlajointpriors}.

\subsection{Model implementation}

We start by loading the data and resetting the binary variable \texttt{sex} to a 0-1 format:
\begin{verbatim}
R> library(frailtyHL)
R> data(kidney)
R> kidney$sex <- kidney$sex - 1   # 0: male (reference)
\end{verbatim}

So, we setup an INLA survival object and fit the frailty model:
\begin{verbatim}
R> m6.frlt <- joint(formSurv = inla.surv(time = time, event = status) ~ sex + (1 | id),
+                   basRisk = "weibullsurv", id = "id", dataSurv = kidney)
\end{verbatim}

Note that the term \texttt{(1 | id)} specifies that frailty is an intercept random effect by \texttt{id} (i.e., patient-specific). Finally, the posterior distribution summary is given by:
\begin{verbatim}
R> summary(m6.frlt)
Survival outcome
                   mean     sd 0.025quant 0.5quant 0.975quant
Weibull (shape)  1.0720 0.0644     0.9591   1.0675     1.2119
Weibull (scale)  0.0205 0.0076     0.0093   0.0192     0.0391
sex             -1.4442 0.3980    -2.2321  -1.4418    -0.6695

Frailty term variance
              mean     sd 0.025quant 0.5quant 0.975quant
IDIntercept 0.4531 0.1932     0.1793   0.4191     0.9295

log marginal-likelihood (integration)    log marginal-likelihood (Gaussian) 
                             -343.211                              -343.211 

Deviance Information Criterion:  665.3281
Widely applicable Bayesian information criterion:  666.2153
Computation time: 0.65 seconds
\end{verbatim}

\subsection{Model interpretation} \label{subsec:FRAILint}

The covariate sex was found to have a statistically significant effect on the time to infections, the hazard ratio can be computed as follows:
\begin{verbatim}
R> summary(m6.frlt, hr=T)$SurvEff[[1]]["sex_S1",]
       exp(mean)     sd 0.025quant 0.5quant 0.975quant
sex_S1    0.2547 0.1022     0.1082   0.2363     0.5066
\end{verbatim}

This suggests that females are associated with a 75\% [49\%, 90\%] reduced susceptibility to infections when accounting for unobserved or latent individual-specific factors captured by the frailty term.
\section{Joint models of longitudinal and survival data} \label{sec:JM1}

\subsection{Motivation: prothro and prothros datasets (from the \texttt{JMbayes} R-package)}

The \texttt{prothro} (longitudinal format) and \texttt{prothros} (survival format) datasets contain 488 liver cirrhosis patients from a placebo-controlled randomised trial, where the longitudinal observations of a biomarker (prothrombin) are recorded.\cite{andersen1993} Each patient has the following variables:
\begin{itemize}
	\item \texttt{id}: patient number.
	\item \texttt{pro}: prothrombin measurements.
	\item \texttt{time}: time points at which the prothrombin measurements were taken.
	\item \texttt{treat}: randomised treatment (placebo or prednisone).
	\item \texttt{Time}: time (in years) from the start of treatment until death or censoring.	
	\item \texttt{death}: censoring indicator (1: if patient is died; 0: otherwise).	
\end{itemize}

41\% of patients received placebo; the number of longitudinal observations (prothrombin measurements) per patient varied between 1 and 7, with 5 being the highest frequency; and the observed survival times had 0.1 years as minimum, 13.4 as maximum, and 2.6 as median, with 40.2\% right-censored.

\subsection{Model specification}

The longitudinal submodel is expressed through a linear mixed-effects specification:
\begin{equation}
y_i(t \mid {\bm \phi}, {\bm b}_{i}, \tau) = \phi_{1} + b_{1i} + (\phi_{2} + b_{2i}) \, t + \phi_{3}\texttt{treat} + \epsilon_i(t \mid \tau), \quad i=1,\ldots,n, \label{eq:JM1}
\end{equation}

\noindent where $y_i(t \mid \cdot)$ represents the log-prothrombin value for patient $i$ at time $t$; $\phi_{1}$ and $\phi_{2}$ are fixed effects for intercept and slope, respectively, with $b_{1i}$ and $b_{2i}$ being the respective individual random effects; $\phi_{3}$ is the regression coefficient for \texttt{treat} covariate; and $\epsilon_i(t \mid \tau)$ is a measurement error for patient $i$ at time $t$. We assume that the individual random effects, ${\bm b_i} = (b_{1i},b_{2i})^{\top}$, given $\Sigma$ follow a joint bivariate normal distribution with zero mean vector and variance-covariance matrix $\Sigma$; and that the errors are conditionally i.i.d. as $\epsilon_i(t \mid \tau) \sim \mbox{Normal}(0,\tau)$, where $\tau$ represents the error precision (defined as one divided by the variance). Random effects and error terms are assumed mutually independent.

The survival submodel is expressed through a PH specification:
\begin{equation}
h(t \mid \lambda, \alpha, \beta, \gamma, {\bm b}_{i}) = h_{0}(t \mid \lambda, \alpha) \exp\left\{\beta\texttt{treat} + \gamma(b_{1i} + b_{2i}t) \right\}, \quad t > 0, \label{eq:JM2}
\end{equation}

\noindent with $h_{0}(t \mid \lambda, \alpha)=\lambda \, \alpha \, t^{\alpha-1}$ specified as a Weibull baseline hazard function, where $\lambda$ and $\alpha$ are the scale and shape parameters, respectively; $\beta$ is the regression coefficient for the \texttt{treat} covariate; and $\gamma$ is an association parameter that measures the strength of the link between both submodels sharing random effects.

Prior distributions for the parameters of both submodels are set according to Section~\ref{subsec:inlajointpriors}.

\subsection{Model implementation}

We start by loading the data:
\begin{verbatim}
R> library(JMbayes)
R> data(prothro)
R> data(prothros)
\end{verbatim}

So, we setup an INLA survival object and fit the joint model:
\begin{verbatim}
R> m7.jm1 <- joint(formSurv = inla.surv(time = Time, event = death) ~ treat, dataSurv = prothros,
+                  formLong = pro ~ time + treat + (1 + time | id), 
+                  dataLong = prothro, basRisk = "weibullsurv", family = "lognormal",
+                  id = "id", timeVar = "time", assoc = "SRE")   
\end{verbatim}

Note that the survival formula specification in the \texttt{joint} function is similar to a PH model. On the other hand, the longitudinal formula specification includes the outcome (\texttt{pro}), the regression structure indicating intercept and slope patient-specific random effects (\texttt{(1 + time | id)}), the family name for sampling distribution (\texttt{family = "lognormal"}, i.e. we fit the outcome on a logarithmic scale), the time variable (\texttt{timeVar = "time"}), and the type of association between the longitudinal and survival submodels (\texttt{assoc = "SRE"}). Specifically, \texttt{SRE} shares the individual deviation from the mean at time $t$ as defined by the random effects, i.e., as in \eqref{eq:JM2}. See \texttt{help(joint)} for other types of association. Finally, the posterior distribution summary is given by:
\begin{verbatim}
R> summary(m7.jm1, sdcor = TRUE)
Longitudinal outcome (lognormal)
                      mean     sd 0.025quant 0.5quant 0.975quant
Intercept_L1        4.2757 0.0215     4.2335   4.2757     4.3178
time_L1            -0.0014 0.0066    -0.0143  -0.0014     0.0115
treatprednisone_L1 -0.0973 0.0305    -0.1572  -0.0973    -0.0374
Res. err. (sd)      0.2577 0.0039     0.2500   0.2577     0.2652

Random effects standard deviation / correlation (L1)
                        mean     sd 0.025quant 0.5quant 0.975quant
Intercept_L1          0.3092 0.0126     0.2888   0.3079     0.3369
time_L1               0.1118 0.0068     0.0982   0.1124     0.1240
Intercept_L1:time_L1 -0.0410 0.0237    -0.0906  -0.0396     0.0013

Survival outcome
                     mean     sd 0.025quant 0.5quant 0.975quant
Weibull (shape)_S1 0.8808 0.0175     0.8455   0.8810     0.9144
Weibull (scale)_S1 0.1990 0.0199     0.1627   0.1980     0.2408
treatprednisone_S1 0.0753 0.1347    -0.1889   0.0753     0.3394

Association longitudinal - survival
             mean     sd 0.025quant 0.5quant 0.975quant
SRE_L1_S1 -2.1562 0.1323    -2.4014   -2.161    -1.8817

log marginal-likelihood (integration)    log marginal-likelihood (Gaussian) 
                            -23339.31                             -23339.31 

Deviance Information Criterion:  9461.311
Widely applicable Bayesian information criterion:  8888.578
Computation time: 8.64 seconds
\end{verbatim}

In order for the variability parameters to be expressed in terms of variance instead of standard deviation, set ``sdcor'' as \texttt{FALSE} or remove it from the summary function call. The \texttt{plot()} function can be applied to any model to return several plots, such as:
\begin{verbatim}
R> plot(m7.jm1, sdcor = FALSE)   # Variance instead of standard deviation
\end{verbatim}

\begin{figure}[ht]
\centering
\includegraphics[scale=0.75]{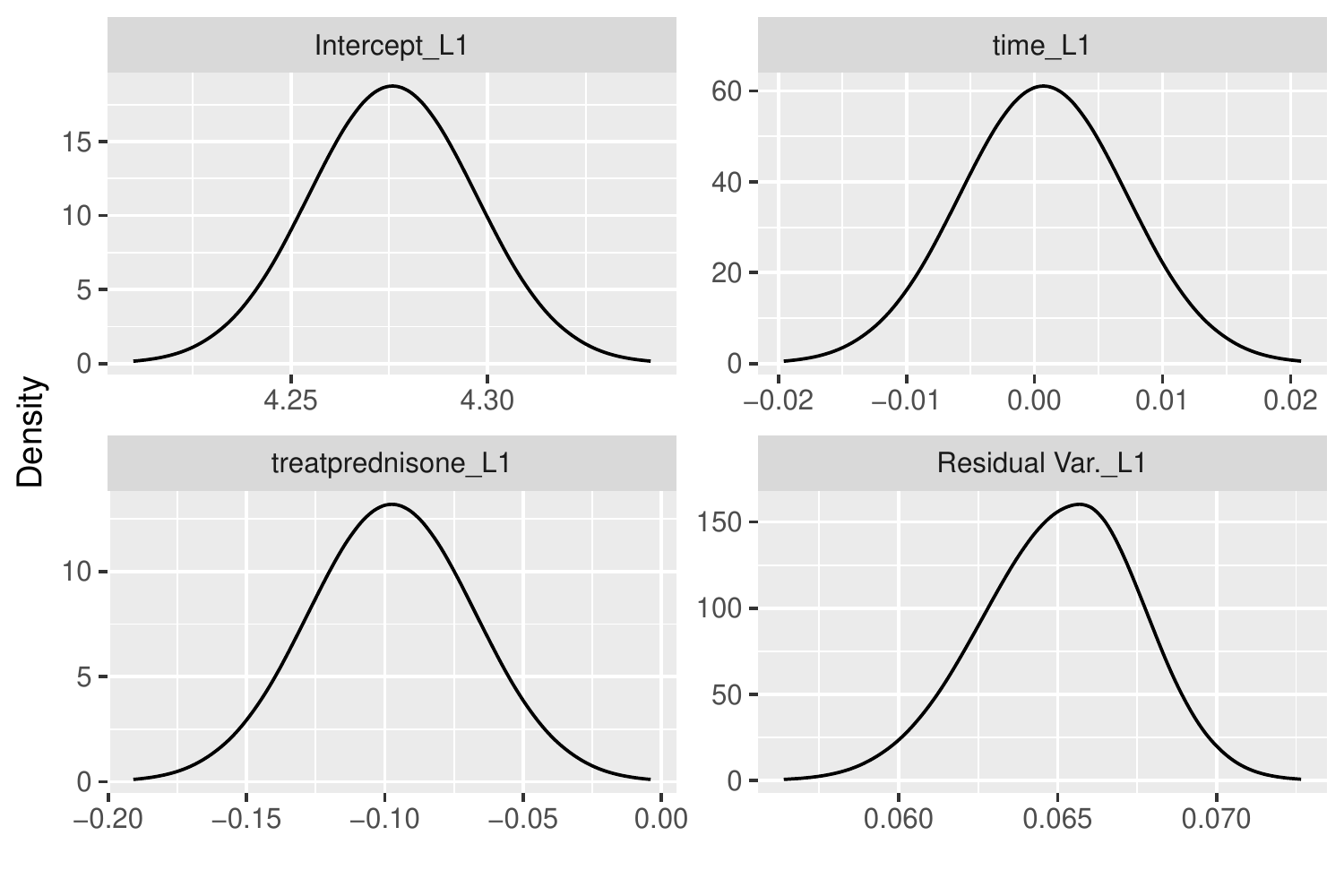}
\caption{Posterior marginal distributions for the fixed effects ($\bm \phi$) and variance of the residual error term of the longitudinal submodel \eqref{eq:JM1}.} \label{fig:longiFixedPost}
\end{figure}
 
\begin{figure}[ht]
\centering
\includegraphics[scale=0.75]{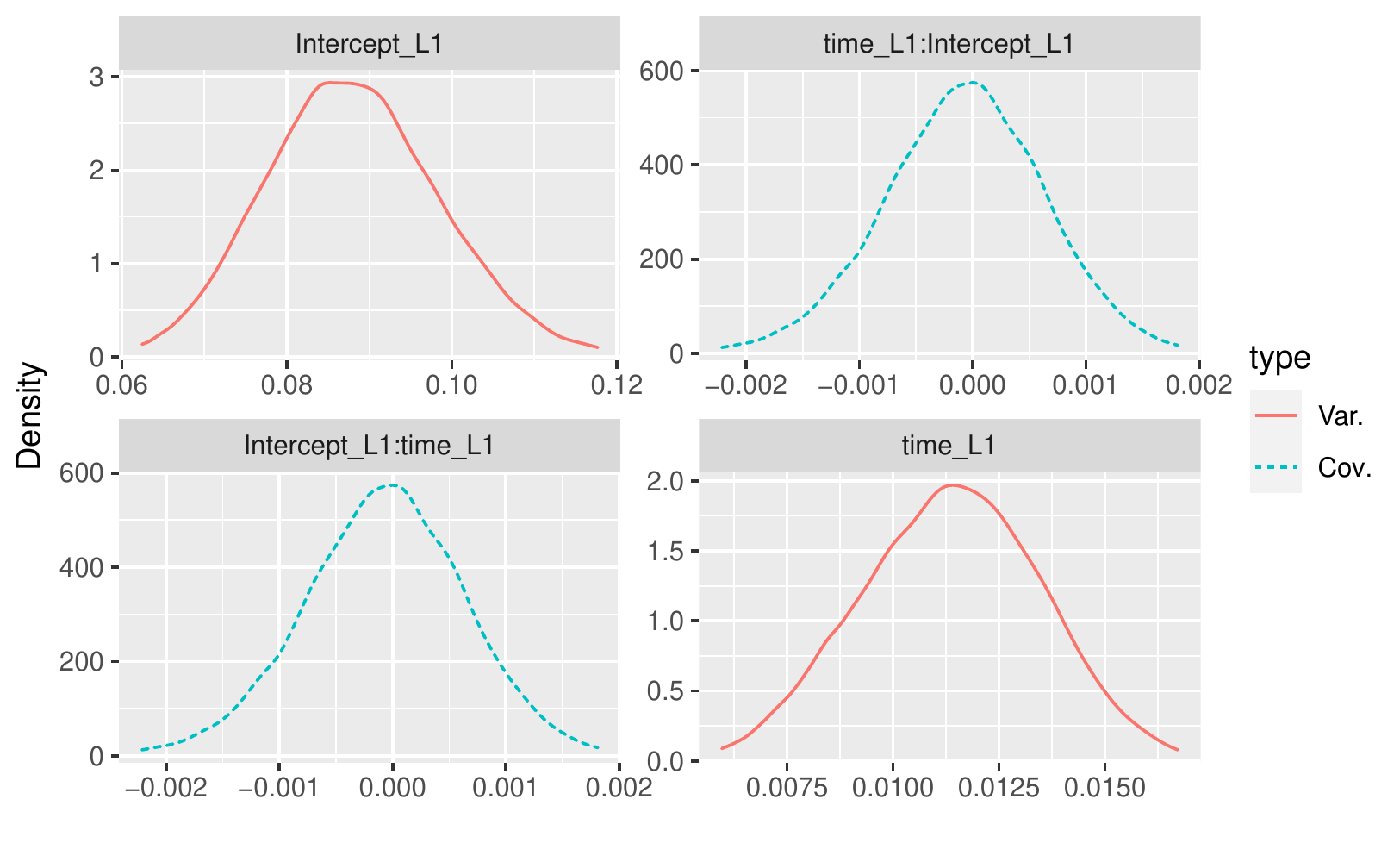}
\caption{Posterior marginal distributions for the variance-covariance matrix ($\Sigma$) of the random effects.} \label{fig:longiRandPost}
\end{figure}

\begin{figure}[ht]
\centering
\includegraphics[scale=0.75]{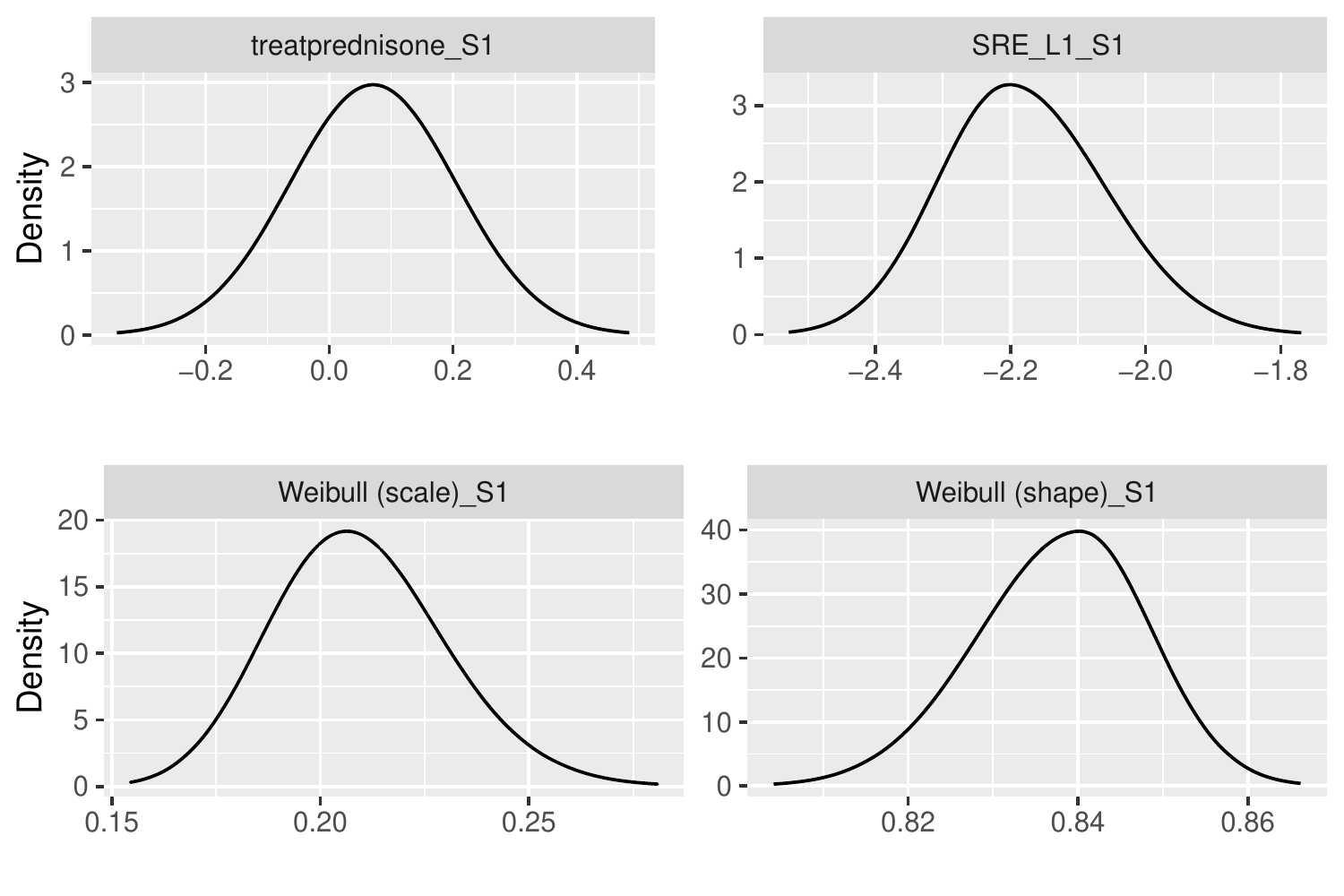}
\caption{Posterior marginal distributions for the fixed effects ($\beta$ and $\gamma$) and Weibull baseline hazard parameters of the survival submodel \eqref{eq:JM2}.} \label{fig:survPost}
\end{figure}

Figure \ref{fig:longiFixedPost} gives the posterior distributions for the the fixed effects ($\bm \phi$) and variance of the residual error term of the longitudinal submodel while Figure \ref{fig:longiRandPost} shows the posterior distribution for variance and covariance terms of the random effects and Figure \ref{fig:survPost} gives the posterior distributions of parameters related to the survival submodel. Visualising posterior distributions can help understanding the uncertainty over parameters compared to standard deviations and quantiles provided by the summary functions as it can clearly display departure from Gaussianity (e.g., heavy tails and skewness).

\subsection{Model interpretation} \label{subsec:JMint}

There are multiple aspects of a joint model that can be interpreted. For example, we can start with the effect of treatment on prothrombin measurements $\phi_3$. In the context of the lognormal distribution, we can obtain a multiplicative effect of covariates by exponentiating the fixed effects value as follows:
\begin{verbatim}
R> inla.zmarginal(inla.tmarginal(function(x) exp(x), m7.jm1$marginals.fixed[["treatprednisone_L1"]]))
      mean         sd quant0.025  quant0.25   quant0.5  quant0.75 quant0.975 
    0.9077     0.0275     0.8547     0.8887     0.9072     0.9261     0.9629 
\end{verbatim}

The treatment prednisone is associated with a statistically significant average decrease of 9\% [3\%, 15\%] in prothrombin level when accounting for individual heterogeneity and informative censoring by death. The association parameter $\hat\gamma$ = -2.16 [-2.41, -1.88] indicates a statistically significant effect of the heterogeneity of the population captured by the random effects in the longitudinal part on the risk of death, meaning that patients with higher/lower prothrombin levels compared to the population average are associated to a reduced/increased risk of death at any given time point, respectively. Finally, the effect of treatment on the risk of death $\hat\beta$ = 0.08 [-0.19, 0.34] is not statistically significant, meaning that there is no evidence for different risk profiles for each treatment line when taking into account individual heterogeneity of the population due to factors not included in the model.
\section{Joint models of longitudinal semicontinuous, recurrent events and terminal event data} \label{sec:JM2}

\subsection{Motivation: colorectalLongi and colorectal datasets (from the \texttt{frailtypack} R-package)}

The \texttt{colorectalLongi} (longitudinal format) and \texttt{colorectal} (survival format) datasets contain a random subsample of 150 patients from the follow-up of the FFCD 2000-05 multicenter phase III clinical trial with metastatic colorectal cancer randomised into two therapeutic strategies: combination and sequential.\cite{ducreux2011} Each patient has the following variables: \\

\noindent \textbf{Only longitudinal format:}
\begin{itemize}
        \item \texttt{year}: time of visit (in years) counted from baseline.
        \item \texttt{tumor.size}: individual longitudinal measurement of transformed (Box-Cox with parameter $0.3$) sums of the longest diameters.
\end{itemize}

\noindent \textbf{Only survival format:}
\begin{itemize}
        \item \texttt{time0}: start of interval (0 or previous recurrence time).
        \item \texttt{time1}: recurrence or censoring time.
        \item \texttt{new.lesions}: appearance of new lesions status (1: new lesions; 0: censored or no event).
        \item \texttt{state}: death indicator (1: dead; 0: alive).
        \item \texttt{gap.time}: intercurrence time or censoring time.
\end{itemize}

\noindent \textbf{Both formats:}
\begin{itemize}
        \item \texttt{id}: patient number.
        \item \texttt{treatment}: to which treatment arm a patient was allocated? (1: sequential (S); 2: combination (C)).
        \item \texttt{age}: age at baseline (1: $<50$ years, 2: $50-69$ years, 3: $>69$ years).
        \item \texttt{who.PS}: WHO performance status at baseline (1: status 0; 2: status 1; 3: status 2).
        \item \texttt{prev.resection}: previous resection of the primate tumour? (1: yes; 0: no).
\end{itemize}

51\% of patients received sequential treatment; the number of longitudinal observations (tumour size measurements) per patient varied between 1 and 18, with 5 being the highest frequency; the proportion of zeros in the longitudinal outcome is 3.8\%. The number of recurrent events (i.e., new lesions) ranges between 0 and 4 per individual with a median of 1, these recurrent events are observed between 0.12 and 2.70 years of follow-up with a median of 0.75. The observed death times had 0.04 years as minimum, 3.84 as maximum, and 0.96 as median, with 19.3\% right-censored.

\subsection{Model specification}

The proposed model is a new model that extends the two-part joint model for a longitudinal semicontinuous biomarker and a terminal event introduced by Rustand et al\cite{rustand2020} and successfully implemented within the INLA framework to overcome the limitations of standard estimation strategies such as Newton-like algorithms or MCMC (i.e., long computation time and convergence issues).\cite{rustand2020bis} Here we include an additional submodel that accounts for recurrent events, linked with all the other submodels through shared random effects. In the context of the application, this new model is able to handle the appearance of new lesions over the follow-up of the clinical trial in addition to the longitudinal semicontinuous measurements of the target tumours measurements and the time to death, which is an important additional piece of information.

The longitudinal submodel is expressed through a two-part mixed-effects specification:
\begin{align}
 Z_{i}(t \mid {\bm \phi}_{1}, b_{1i}) &\sim \mbox{Bernoulli}\big(\eta_{i}(t \mid {\bm \phi}_{1}, b_{1i})\big), \quad i=1,\ldots,n, \label{eq:JMbin} \\ 
\mbox{logit}\big(\eta_{i}(t \mid {\bm \phi}_{1}, b_{1i})\big) &= \phi_{11} + b_{1i} + \phi_{12}t + \phi_{13}\texttt{treatment} + \phi_{14} \times t \times \texttt{treatment}, \label{eq:JMbin2} \\
 \log\left(y_{i}(t \mid {\bm \phi}_{2}, {\bm b}_{2i}, \tau, y_i(t)>0)\right) &= \phi_{21} + b_{21i} + (\phi_{22} + b_{22i}) \, t + \phi_{23}\texttt{treatment} + \nonumber \\
  & \quad \,\, \phi_{24} \times t \times \texttt{treatment} + \epsilon_{i}(t \mid \tau), \label{eq:JMcon}
\end{align}

\noindent where $Z_{i}(t \mid \cdot)$ is a binary variable defined as $0$ if $y_i(t \mid \cdot)=0$ and $1$ otherwise, with $y_i(t \mid \cdot)$ being the tumour size for patient $i$ at time $t$; ${\bm\phi}_{1}$ and ${\bm\phi}_{2}$ are fixed effects parameters of binary \eqref{eq:JMbin} and semicontinuous \eqref{eq:JMcon} parts, including \texttt{treatment}, time (\texttt{year}) and their interaction; and $\epsilon_i(t \mid \tau)$ is a normally distributed error term for patient $i$ at time $t$. For the binary submodel, we include an intercept random effect, $b_{1i}$ while for the semicontinuous submodel, we consider intercept and slope random effects, ${\bm b_{2i}} = (b_{21i},b_{22i})^{\top}$. The random effects follow a joint multivariate normal distribution with zero mean vector and variance-covariance matrix $\Sigma$, thus accounting for correlation between repeated measurements of an individual as well as the correlation between the binary and continuous parts. Random effects and error terms are assumed mutually independent.

The recurrent events corresponding to the appearance of new lesions are modelled with a frailty model \eqref{eq:jmrec} and the survival times are modelled with a PH model \eqref{eq:jmter}. The frailty term is shared in the PH submodel, it is normally distributed and assumed independent from random effects in the longitudinal parts. In addition, the random effects from the two-part mixed-effects model are shared in both the frailty model and the PH model:
\begin{align}
	h_{1}(t \mid \lambda_{1}, {\beta}_{1}, \bm\gamma_1, b_{1i}, {\bm b}_{2i}, b_{3i}) &= h_{01}(t \mid \lambda_{1})\exp\big\{\beta_{1}\texttt{treatment}_i + \nonumber \\ 
    & \qquad \qquad \qquad \qquad \gamma_{11}b_{1i}+\gamma_{12}b_{21i}+\gamma_{13}b_{22i} + b_{3i}\big\}, \quad t > 0,  \label{eq:jmrec} \\
	h_{2}(t \mid \lambda_{2}, {\beta}_{2}, \bm\gamma_2, b_{1i}, {\bm b}_{2i}, b_{3i}) &= h_{02}(t \mid \lambda_{2})\exp\big\{\beta_{2}\texttt{treatment}_i + \nonumber \\
    & \qquad \qquad \qquad \qquad \gamma_{21}b_{1i}+\gamma_{22}b_{21i}+\gamma_{23}b_{22i} + \gamma_{33}b_{3i}\big\}, \quad t > 0.  \label{eq:jmter}
\end{align}

The baseline hazard function of both survival submodels is approximated with smooth splines corresponding to a second-order random walk prior with precision parameters $\lambda_{1}$ and $\lambda_{2}$; $\beta_1$ and $\beta_2$ are the regression coefficient for the \texttt{treatment} covariate in the two survival submodels \eqref{eq:jmrec} and \eqref{eq:jmter}, respectively; and $\gamma$'s are association parameters that scale the shared random effects and measure the strength of the link between the submodels. 

Prior distributions for the parameters of both submodels are set according to Section~\ref{subsec:inlajointpriors}.

\subsection{Model implementation}

We start by loading the data:
\begin{verbatim}
R> library(frailtypack)
R> data(colorectal)
R> data(colorectalLongi)
\end{verbatim}

Then, we reverse the Box-Cox transformation to have tumour size on the original scale, select only positive sizes of the tumour, and create the binary outcome which indicates the presence of tumour:
\begin{verbatim}
R> colorectalLongi$y <- round((colorectalLongi$tumor.size*0.3+1)^(1/0.3), 5)
R> colorectalLongiPositive <- colorectalLongi[colorectalLongi$y > 0,]
R> colorectalLongi$z <- ifelse(colorectalLongi$y == 0, 0, 1)
\end{verbatim}

So, we extract terminal event data for submodel \eqref{eq:jmter}, set INLA survival objects for recurrent and terminal events, and fit the joint model:
\begin{verbatim}
R> colorectalSurv <- subset(colorectal, new.lesions == 0)
R> m8.jm2 <- joint(formSurv = list(inla.surv(time = time1, truncation = time0, 
                                             event = new.lesions) ~ treatment + (1 | id),
+                                  inla.surv(time = time1,  event = state)~ treatment), 
+                  formLong = list(z ~ year * treatment + (1 | id), 
+                                  y ~ year * treatment + (1 + year | id)), 
+                  dataSurv = list(colorectal, colorectalSurv), 
+                  dataLong = list(colorectalLongi, colorectalLongiPositive),
+                  basRisk = c("rw2","rw2"), assocSurv = TRUE, 
+                  timeVar = "year", family = c("binomial","lognormal"), id = "id", 
+                  corLong = TRUE, assoc = list(c("SRE_ind","SRE_ind"), c("SRE_ind","SRE_ind"))) 
\end{verbatim}

The association between the longitudinal and survival submodels (\texttt{assoc = "SRE\_ind"}) indicates that each random effect is shared independently in the survival submodels, as described in \eqref{eq:jmrec} and \eqref{eq:jmter} (as opposed to the individual deviation at time $t$ defined by the random effects, as illustrated in Section \ref{sec:JM1}). This association is decomposed in a list of two vectors with two association parameters in each that corresponds to the two longitudinal submodels \eqref{eq:JMbin} and \eqref{eq:JMcon} sharing into the two survival submodels \eqref{eq:jmrec} and \eqref{eq:jmter}, respectively (see \texttt{help(joint)} for other types of association). The argument \texttt{corLong = TRUE} indicates that the random effects across the two longitudinal submodels are correlated (setting this parameter to \texttt{FALSE} would make the random intercept in \eqref{eq:JMbin} independent from the random intercept and slope from \eqref{eq:JMcon}). The argument \texttt{assocSurv = TRUE} indicates that the frailty term from the recurrent events submodel is shared in the terminal event submodel. Finally, the posterior distribution summary is given by:
\begin{verbatim}
R> summary(m8.jm2)
Longitudinal outcome (L1, binomial)
                      mean     sd 0.025quant 0.5quant 0.975quant
Intercept_L1        6.9695 1.0946     4.8241   6.9695     9.1148
year_L1             0.2716 1.0366    -1.7600   0.2716     2.3033
treatmentC_L1      -2.0468 1.1744    -4.3485  -2.0468     0.2550
year:treatmentC_L1 -1.0604 1.0659    -3.1495  -1.0604     1.0287

Longitudinal outcome (L2, lognormal)
                        mean     sd 0.025quant 0.5quant 0.975quant
Intercept_L2          2.1274 0.0845     1.9617   2.1274     2.2931
year_L2              -0.1432 0.0780    -0.2961  -0.1432     0.0098
treatmentC_L2         0.0539 0.1212    -0.1838   0.0539     0.2915
year:treatmentC_L2   -0.3157 0.1116    -0.5345  -0.3157    -0.0969
Res. err. (variance)  0.1647 0.0093     0.1472   0.1644     0.1838

Random effects variance-covariance
                             mean     sd 0.025quant 0.5quant 0.975quant
Intercept_L1               2.6064 1.5135     0.7542   2.2540     6.6371
Intercept_L2               0.4888 0.0664     0.3738   0.4836     0.6300
year_L2                    0.1854 0.0503     0.1025   0.1798     0.2978
Intercept_L1:Intercept_L2  0.7259 0.2691     0.2605   0.6980     1.3347
Intercept_L1:year_L2      -0.1882 0.1793    -0.6226  -0.1617     0.0908
Intercept_L2:year_L2      -0.0914 0.0498    -0.2103  -0.0848    -0.0131

Survival outcome (S1)
                               mean     sd 0.025quant 0.5quant 0.975quant
Baseline risk (variance)_S1  0.1343 0.0896     0.0315   0.1117     0.3730
treatmentC_S1               -0.2842 0.2129    -0.7015  -0.2842     0.1332

Frailty term variance (S1)
                 mean     sd 0.025quant 0.5quant 0.975quant
IDIntercept_S1 0.5228 0.2901     0.1699   0.4534     1.2857

Survival outcome (S2)
                               mean     sd 0.025quant 0.5quant 0.975quant
Baseline risk (variance)_S2  0.0590 0.0554     0.0071   0.0425     0.2122
treatmentC_S2               -0.1457 0.3179    -0.7688  -0.1457     0.4774

Association longitudinal - survival
                       mean     sd 0.025quant 0.5quant 0.975quant
SRE_Intercept_L1_S1  0.2230 0.2487    -0.2812   0.2279     0.6977
SRE_Intercept_L1_S2  0.5471 0.4352    -0.3324   0.5547     1.3808
SRE_Intercept_L2_S1 -0.1158 0.4017    -0.8841  -0.1233     0.6972
SRE_year_L2_S1      -0.0877 0.3444    -0.7570  -0.0907     0.5991
SRE_Intercept_L2_S2  0.3320 0.6159    -0.8565   0.3239     1.5685
SRE_year_L2_S2      -0.4435 0.5100    -1.4632  -0.4382     0.5448

Association survival - survival
                    mean     sd 0.025quant 0.5quant 0.975quant
IDIntercept_S1_S2 1.5642 0.3673     0.8232   1.5702     2.2694

log marginal-likelihood (integration)    log marginal-likelihood (Gaussian) 
                            -3534.745                             -3518.836 

Deviance Information Criterion:  6436.598
Widely applicable Bayesian information criterion:  6458.738
Computation time: 18.08 seconds
\end{verbatim}

\begin{figure}[ht]
\centering
\includegraphics[scale=0.7]{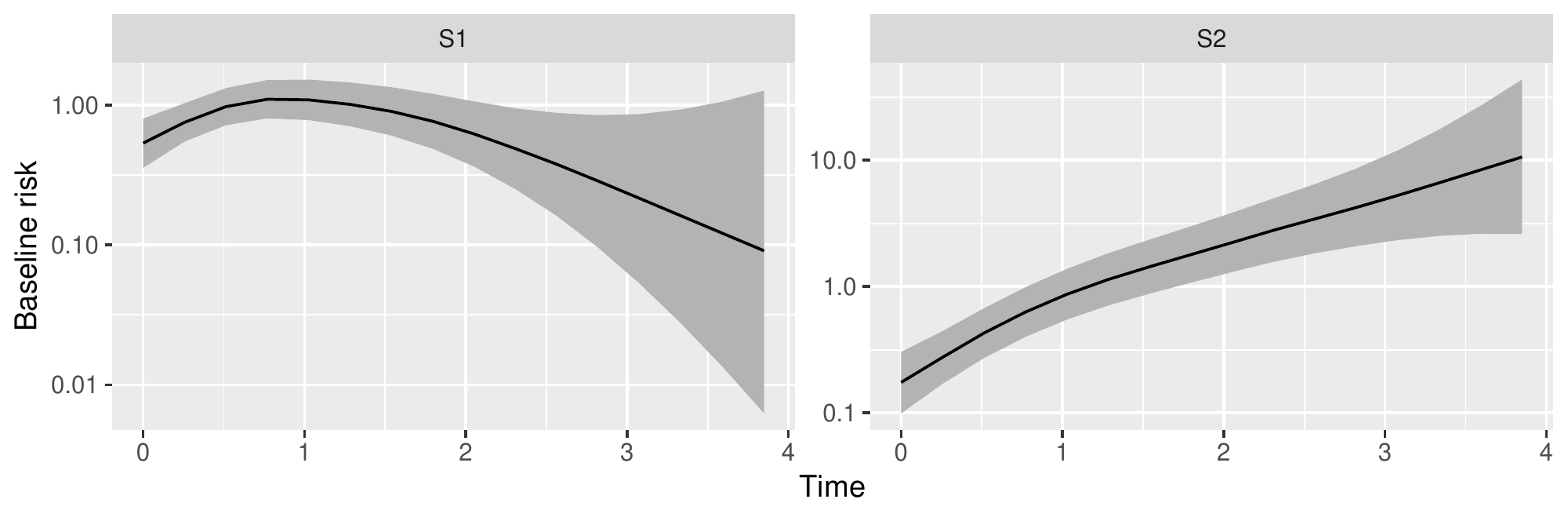}
\caption{Baseline hazard curves from joint model \eqref{eq:JMbin} fitted by smooth splines with 15 equidistant knots.} \label{fig:BH}
\end{figure}

The recurrent events are set to use calendar times, i.e., the beginning of the ``at-risk'' period for the second recurrent event starts after the time of occurrence of the first event. However, it is possible to use gap times in order to have a clock-reset after each recurrent event:
\begin{verbatim}
R> surv.obj.jm2.recu <- inla.surv(time = colorectal$gap.time, event = colorectal$new.lesions)
\end{verbatim}

Baseline hazard curves can be plotted (in log10 scale) using \texttt{ggplot2} R-package. Figure \ref{fig:BH} displays the baseline hazards curves for each of the two competing risks. The baseline hazard for the recurrent events submodel shows an initial increase of the hazard followed by a decrease, clearly illustrating the usefulness of using Bayesian smooth splines instead of parametric distributions while the second curve corresponding to the baseline hazard of death shows a constant increase that could alternatively be captured with some parametric distribution such as Weibull.
\begin{verbatim}
R> plot(m8.jm2)$Baseline + scale_y_log10()
\end{verbatim}

\subsection{Model interpretation} \label{subsec:JM2int}
The first part of the model displayed in the summary is the longitudinal binary part that evaluates the effect of covariates on the probability of positive tumour size. In the dataset, only a few zeros are observed, thus leading to a high probability of positive value. For example, we can compute the probability of positive tumour size 3 years after randomisation conditional on treatment (credible intervals follow trivially from sampling):
\begin{verbatim}
R> inv.logit(m8.jm2$summary.fixed["Intercept_L1", 1] + 3*m8.jm2$summary.fixed["year_L1", 1])
0.9996
R> inv.logit(m8.jm2$summary.fixed["Intercept_L1", 1] + 3*m8.jm2$summary.fixed["year_L1", 1] +
+            m8.jm2$summary.fixed["treatmentC_L1", 1] +
+            3*m8.jm2$summary.fixed["year.X.treatmentC_L1", 1])
0.9280
\end{verbatim}

This suggests that the probability of positive tumour size 3 years after randomization for an average patient is 0\% for sequential treatment while it is 7\% for the combination treatment, accounting for informative censoring related to the appearance of new lesions and death. Note that due to the non-linear link function, this interpretation is subject-specific and corresponds to individuals with random effects equal to zero (i.e., average patients). From the second part of the model, we can compute the multiplicative effect of treatment on the average tumour size evolution over time, conditional on observing a positive tumour size:
\begin{verbatim}
R> inla.zmarginal(inla.tmarginal(function(x) exp(x), m8.jm2$marginals.fixed[[10]]))
      mean         sd quant0.025  quant0.25   quant0.5  quant0.75 quant0.975 
    0.7338     0.0816     0.5864     0.6763     0.7292     0.7861     0.8067 
\end{verbatim}

Patients with positive tumour size receiving the combination treatment are associated with a statistically significant average decrease of 27\% [19\%, 42\%] in tumour size per year of follow-up compared to patients receiving sequential treatment, accounting for informative censoring related to the appearance of new lesions and death. The association parameters between the longitudinal markers and the recurrent and terminal event submodels give the effect of the heterogeneity captured by each random effect independently shared in the survival submodels, this can be useful to account for individual heterogeneity due to unobserved confounders in survival models. As opposed to the previous joint model (see Section \ref{sec:JM1}) where the linear combination of the random effects was shared, here we share and scale each random effect separately, thus offering more flexibility in the association. The frailty term in the recurrent event submodel captures some additional individual heterogeneity due to unobserved factors in the risk of the appearance of new lesions. This frailty term is shared and scaled in the survival model, the scaling parameter is $\hat\gamma_{33}$ = 1.56 [0.82, 2.27]. While the association of the random effects from the longitudinal submodels with the survival submodels is not significantly different from zero, this frailty term effect on the risk of death is statistically significant, indicating that the individual heterogeneity in the risk of appearance of new lesions is predictive of the risk of death (i.e., higher risk of new lesions lead to higher risk of death).

\section{Model checks and prior sensitivity analysis} \label{sec:PS}
In Bayesian inference, a model that is ill-defined or non-identifiable due to a lack of data or a flat likelihood will not fail as it would in the frequentist inference framework because when there is no information to estimate a parameter, the posterior distribution should be driven by the priors. With MCMC, such models are often associated to slow convergence properties while with \texttt{INLAjoint}, the computation time is not affected. Therefore it is important to be able to identify these models. Several warnings are included in \texttt{INLAjoint} to inform the user about potentially ill-defined models, these warnings are triggered under specific circumstances (highly correlated hyperparameters, abnormal hyperparameters skewness correction, number of random effects larger than number of data points) and can help the user to identify potential issues with the model. Additionally, while MCMC-like convergence diagnostics does not apply to the INLA methodology (because INLA is based on numerical approximations to the posterior densities instead of MCMC samples), all the traditional Bayesian model checks holds. We implemented a method to visualise the model sensitivity to priors specification. We illustrate in this section how to perform prior sensitivity analysis with INLAjoint by looking at a simple example. Consider the logistic submodel (i.e., binary part) from the last example defined by Equation \eqref{eq:JMbin} of Section \ref{sec:JM2}. We can fit this model with a specific prior distribution by specifying priors in the control options as follows:
\begin{verbatim}
R> m8.sen <- joint(formLong = z ~ year * treatment + (1 | id),
+                  dataLong = colorectalLongi, timeVar="year",
+                  family = c("binomial"), id = "id",
+                  control=list(priorFixed=list(
+                    mean=0, prec=0.001, mean.intercept=0, prec.intercept=0.001)))
R> print(summary(m8.sen)$FixedEff[[1]], digits=2)
                     mean   sd 0.025quant 0.5quant 0.975quant
Intercept_L1        5.944 1.08        3.8    5.944      8.063
year_L1            -0.065 0.99       -2.0   -0.065      1.876
treatmentC_L1      -2.127 1.13       -4.3   -2.127      0.094
year:treatmentC_L1 -1.165 1.01       -3.2   -1.165      0.823
\end{verbatim}

Here, the fixed effects are assigned a Gaussian prior with mean 0 and variance 1000 (instead of the default 100), note that the prior for the intercept is defined separately through the mean.intercept and prec.intercept control arguments. We can display the prior versus posterior distributions for these fixed effects by calling the plot function with the argument priors=TRUE:

\begin{verbatim}
R> plot(m8.sen, priors=TRUE)$Outcomes$L1
\end{verbatim}

\begin{figure}[ht]
\centering
\includegraphics[scale=0.7]{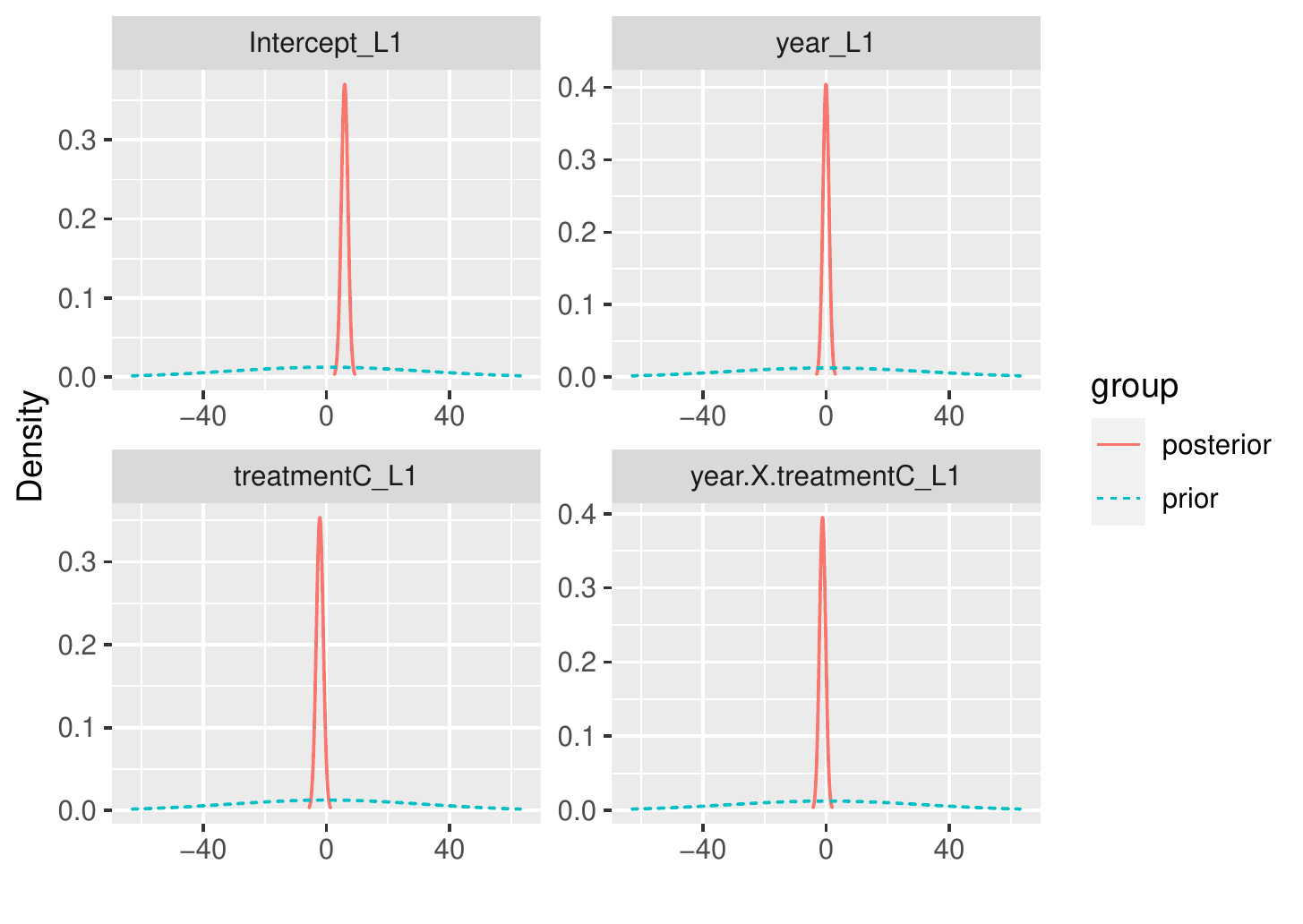}
\caption{Prior and posterior distributions for the fixed effects of the identifiable model described in Section \ref{sec:PS} with Gaussian priors with mean 0 and variance 1000 for all the fixed effects.}\label{fig:PvP1}
\end{figure}

In Figure \ref{fig:PvP1}, the priors are flat while the posteriors are peaked, indicating that posteriors are driven by the information from the data. Since the model is identifiable, changing the prior to another reasonable distribution will have a limited impact on the posterior distributions:
\begin{verbatim}
R> m8.sen2 <- joint(formLong = z ~ year * treatment + (1 | id),
+                  dataLong = colorectalLongi, timeVar="year",
+                  family = c("binomial"), id = "id",
+                  control=list(priorFixed=list(
+                    mean=10, prec=0.01, mean.intercept=10, prec.intercept=1)))
R> print(summary(m8.sen2)$FixedEff[[1]], digits=2)
                     mean   sd 0.025quant 0.5quant 0.975quant
Intercept_L1        5.848 1.04        3.8    5.848       7.89
year_L1            -0.011 0.98       -1.9   -0.011       1.91
treatmentC_L1      -2.024 1.09       -4.2   -2.024       0.12
year:treatmentC_L1 -1.221 1.00       -3.2   -1.221       0.74
\end{verbatim}
\begin{figure}[ht]
\centering
\includegraphics[scale=0.7]{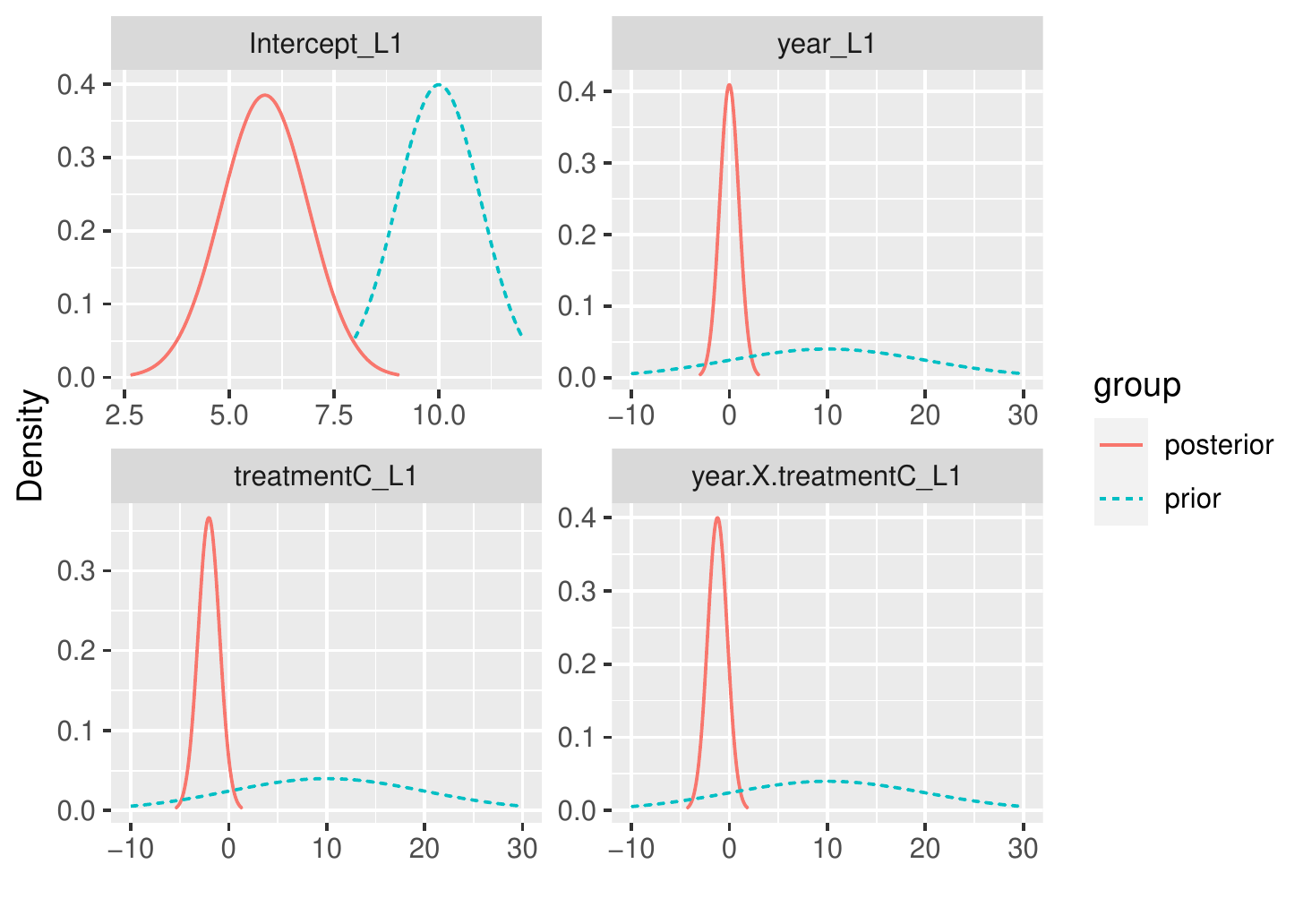}
\caption{Prior and posterior distributions for the fixed effects of the identifiable model described in Section \ref{sec:PS} with Gaussian priors with mean 10 and variance 1 for the fixed intercept, mean 10 and variance 100 for the other fixed effects.}\label{fig:PvP2}
\end{figure}

Figure \ref{fig:PvP2} shows how the posterior distributions are almost unchanged despite a very different prior specification as the posteriors are driven by the data information. Now we can change the data so that the model is not identifiable, for example we can only keep longitudinal measurements at time 0 so that there is no information to identify slope-related parameters, then we can visualise the prior vs. posterior distributions again:
\begin{verbatim}
R> colorectalLongi2 <- colorectalLongi[colorectalLongi$year==0,]
R> m8.sen3 <- joint(formLong = z ~ year * treatment + (1 | id),
+                  dataLong = colorectalLongi2, timeVar="year",
+                  family = c("binomial"), id = "id")
R> plot(m8.sen3, priors=TRUE)$Outcomes$L1
R> print(summary(m8.sen3)$FixedEff[[1]], digits=2)
                      mean   sd 0.025quant 0.5quant 0.975quant
Intercept_L1       1.3e+01  3.7        6.0  1.3e+01         20
year_L1            8.8e-17 10.0      -19.6  1.7e-14         20
treatmentC_L1      6.9e+00  6.8       -6.5  6.9e+00         20
year:treatmentC_L1 8.8e-17 10.0      -19.6  1.7e-14         20
\end{verbatim}
\begin{figure}[ht]
\centering
\includegraphics[scale=0.7]{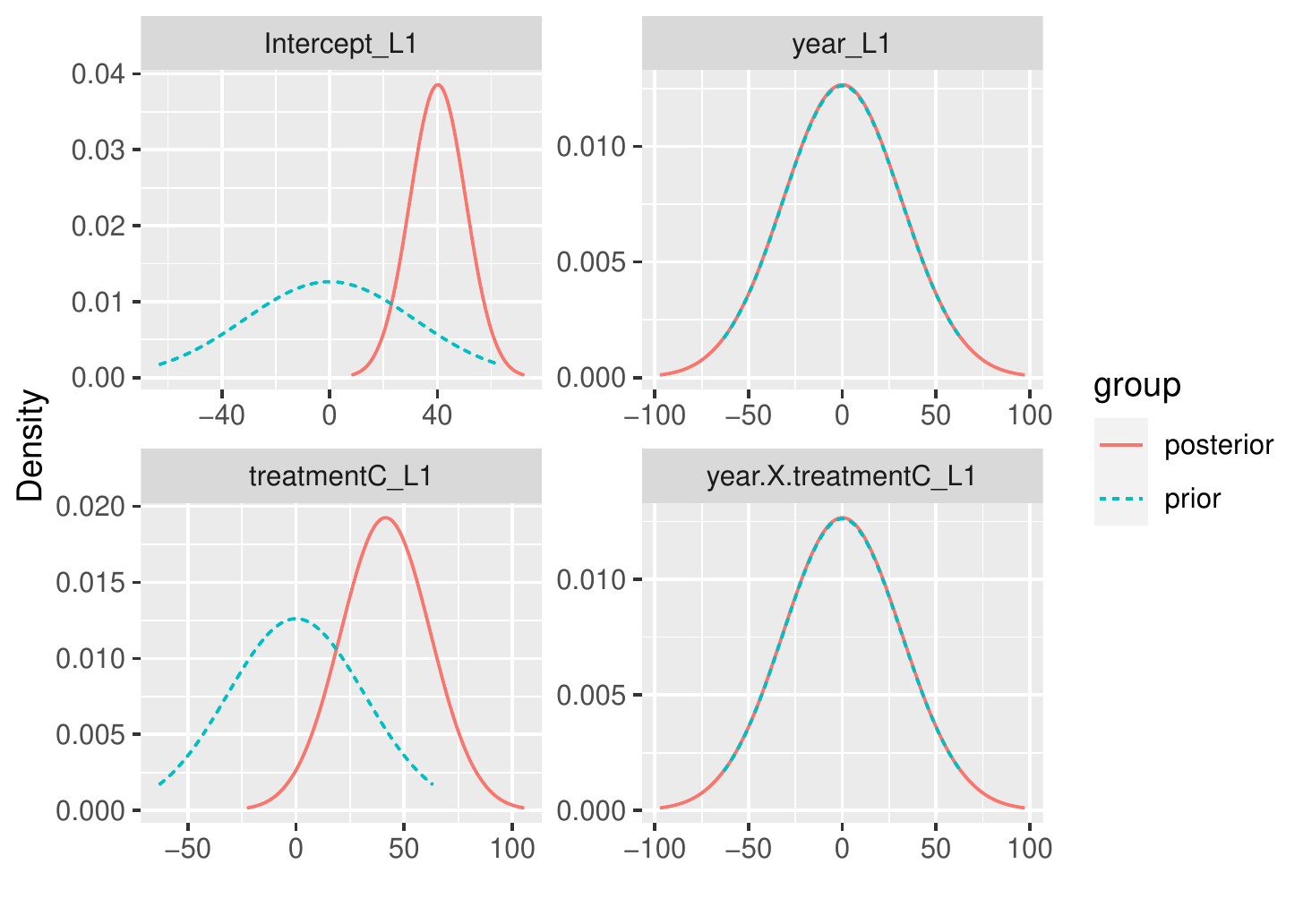}
\caption{Prior and posterior distributions for the fixed effects of the non-identifiable model described in Section \ref{sec:PS}.}\label{fig:PvP3}
\end{figure}

It is clear from Figure \ref{fig:PvP3} that posteriors are now matching priors for the two slope-related parameters, showing a lack of information in the data to estimate the slope-related parameters. Here we used the default prior (i.e., Gaussian with mean 0 and variance 100) but regardless of the prior choice, the posterior will match them as the data is not informative and the model is not identifiable. Evaluating the sensitivity of posterior distributions assuming different priors can help to assess the model stability and reliability. Another example of prior specification and prior sensitivity analysis has been proposed in the supplementary materials of Rustand et al.\cite{rustand2022}, where the illustration is given for the correlation parameter between two random effects of a longitudinal mixed effects model.
\section{Comparison of the fitted models with MCMC} \label{sec:CMP}

\begin{figure}[p!]
\centering
\includegraphics[scale=0.78]{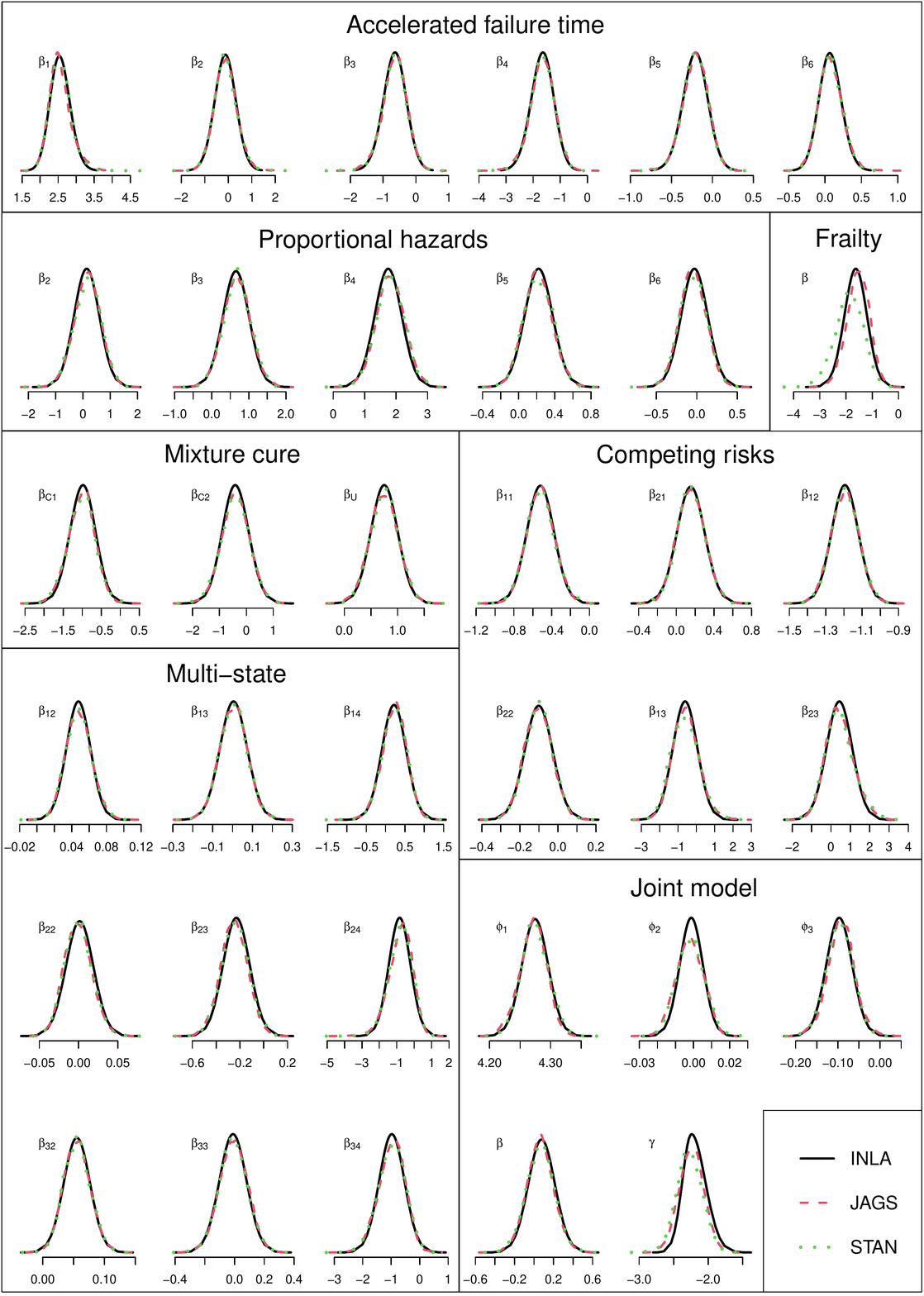}
\caption{Marginal posterior densities for all the fixed effects of the first 7 models fitted with INLA, JAGS, and STAN. Corresponding computation times are provided in Table \ref{CompTime}}
\label{PostDens}
\end{figure}

Comparing the model fit between INLA and MCMC is an important step in assessing the computational efficiency and reliability of our approach. In this paper, we have the opportunity to make such a comparison, as a previous work\cite{alvares2021} fitted all the models (with the exception of the last joint model) using the \texttt{JAGS} software \citep{plummer2003jags} using the R package \texttt{rjags} \citep{rjags23}. Additionally, we fitted these models with \texttt{STAN} \citep{carpenter2017stan} using the R package \texttt{rstan} \citep{rstan24}, another popular implementation of MCMC available in R that relies on the Hamiltonian Monte Carlo method. The last joint model (described in Section \ref{sec:JM2}) is not included in the comparison because writing the code for this model with MCMC becomes more complex as there is no readily available user-friendly package or prior implementation of this model in the literature. This complex joint model aims at illustrating how \texttt{INLAjoint} allows to build more complex models with simple code syntax. We used \texttt{INLA} version 24.02.09, \texttt{INLAjoint} version 24.2.4, \texttt{rjags} version 4-15 with 4 chains, 5000 iterations per chain including 1000 burn-in iterations and thinning of 1 for all models and \texttt{rstan} version 2.32.5 with default options, i.e., 4 chains, 2000 iterations per chain including 1000 burn-in iterations and thinning of 1. Figure \ref{PostDens} displays the posterior marginals for all the fixed effects in each model in order to assess the agreement between the INLA and MCMC results. As expected, these posterior distributions exhibit a high degree of concordance, indicating that INLA and both MCMC methods yield consistent results. This further underscores the reliability of INLA as an alternative to MCMC, particularly when computational efficiency is of utmost importance. Computation time is an essential consideration in Bayesian modelling, especially when dealing with extensive datasets or complex models. To illustrate the computational efficiency of INLA, we detailed in Table \ref{CompTime} the time required for fitting all the models using both INLA and MCMC. This comparison highlights the substantial time savings achievable with INLA. For example, for the first joint model, INLA delivered results within 12 seconds, whereas MCMC required more than 15 minutes with \texttt{STAN} and 29 minutes with \texttt{JAGS} to converge to the same posterior distributions (see Figure \ref{PostDens}).

\begin{table}[ht]
\centering
\begin{tabular}{cccccccc}
  \hline
 & AFT & PH & Mixture cure & Competing risk & Multi-state & Frailty & Joint model 1 \\ 
  \hline
INLA & 0.63 & 0.62 & 0.60 & 1.35 & 0.80 & 0.59 & 12.03 \\ 
  JAGS & 24.25 & 14.95 & 61.55 & 1271.93 & 154.81 & 4.70 & 1753.70 \\ 
  STAN & 67.36 & 66.00 & 74.24 & 349.50 & 94.32 & 72.89 & 948.50 \\ 
   \hline
\end{tabular}
\caption{Computation time (in seconds) for the first 7 models with INLA, JAGS and STAN.}
\label{CompTime}
\end{table}

\section{Conclusions} \label{sec:conclusions}

We have revisited several of the main survival models using integrated nested Laplace approximations (INLA) to perform approximate Bayesian inference. Specifically, we have implemented accelerated failure time, proportional hazard, mixture cure, competing risks, multi-state, frailty, and two joint model specifications for longitudinal and time-to-event data. Of particular note is the last joint model, a novel addition to the literature, which combines a two-part mixed-effects model for a longitudinal semicontinuous outcome, a frailty model for recurrent events, and a proportional hazards model for a terminal event, all linked through shared random effects. This illustrates the flexibility of \texttt{INLAjoint}, where each model component is considered as a building block and thus allows the combination of different components to construct various models that satisfy clinicians' and health researchers' needs. 

In conclusion, our work demonstrates that Bayesian inference with INLA, and specifically the \texttt{INLAjoint} R-package, offers a user-friendly and flexible approach for fitting various survival models. It simplifies the implementation of complex models, requiring considerably fewer lines of code compared to traditional methods such as JAGS or STAN. Its efficiency, both in terms of coding simplicity and computation time, makes it a compelling alternative to MCMC methods. Recently\citep{niekerk2022}, INLA was shown to perform very well for survival analysis in the case of big data (a Cox model with 100 000 patients was fitted in 22.5 seconds). Future research for big data and joint models using INLA is ongoing.
We hope this tutorial has convinced the reader that Bayesian inference with INLA presents an efficient approach to survival analysis.

\bibliography{refs}
\bibliographystyle{unsrt}

\end{document}